\documentclass[amsmath,amssymb,aps,prd,11pt,tightenlines,superscriptaddress,
nofootinbib,preprintnumbers,notitlepage]{revtex4-1}
\usepackage{multirow}
\usepackage{graphicx}
\usepackage{tabularx}
\usepackage{slashed}
\usepackage{url}
\usepackage{hyperref}
\usepackage[maxfloats=100]{morefloats}
\usepackage{color,xcolor}
\usepackage{booktabs}
\usepackage[normalem]{ulem}

\makeatletter
\def\l@subsubsection#1#2{}
\makeatother

\newcommand{\lag}{\ensuremath{\mathcal{L}}}

\newcommand{\qqqquad}{\qquad \qquad \qquad}

\newcommand{\itev}{{\ensuremath\rm TeV^{-1}}}

\def\slashchar#1{\setbox0=\hbox{$#1$}           
   \dimen0=\wd0                                 
   \setbox1=\hbox{/} \dimen1=\wd1               
   \ifdim\dimen0>\dimen1                        
      \rlap{\hbox to \dimen0{\hfil/\hfil}}      
      #1                                        
   \else                                        
      \rlap{\hbox to \dimen1{\hfil$#1$\hfil}}   
      /                                         
   \fi}

\def\eg{{e.\,g.}\ }
\def\ie{{i.\,e.}\ }

\AtBeginDocument{
  \heavyrulewidth=.08em
  \lightrulewidth=.05em
  \cmidrulewidth=.03em
  \belowrulesep=.65ex
  \belowbottomsep=0pt
  \aboverulesep=.4ex
  \abovetopsep=0pt
  \cmidrulesep=\doublerulesep
  \cmidrulekern=.5em
  \defaultaddspace= .5em
  \setlength{\tabcolsep}{0.5em}
}

\begin{document}


\title{Learning from the New Higgs-like Scalar before It Vanishes} 

\author{Martin Bauer}
\affiliation{Institut f\"ur Theoretische Physik, Universit\"at Heidelberg, Germany}

\author{Anja Butter}
\affiliation{Institut f\"ur Theoretische Physik, Universit\"at Heidelberg, Germany}

\author{J.~Gonzalez--Fraile}
\affiliation{Institut f\"ur Theoretische Physik, Universit\"at Heidelberg, Germany}

\author{Tilman Plehn}
\affiliation{Institut f\"ur Theoretische Physik, Universit\"at Heidelberg, Germany}

\author{Michael Rauch}
\affiliation{Institute for Theoretical Physics, Karlsruhe Institute of Technology, Germany}


\begin{abstract} 
  Motivated by a di-photon anomaly observed by ATLAS and CMS we
  develop an \textsc{SFitter} analysis for a combined electroweak-Higgs
  sector, and a scalar portal at the
  LHC. The theoretical description is based on the linear effective
  Lagrangian for the Higgs and gauge fields, combined with an additional singlet scalar.
  The key feature is the extraction of reliable
  information on the portal structure of the combined scalar
  potential. For the specific di-photon anomaly we find that the new
  state might well form such a Higgs portal. To
  obtain more conclusive results we define and test the connection of the
  Wilson coefficients in the Higgs and heavy scalar sectors, as
  suggested by a portal setup.
\end{abstract}

\maketitle
\tableofcontents

\clearpage

\section{Introduction}
\label{sec:intro}

The discovery of a light Higgs boson~\cite{higgs,discovery} has opened
a major new avenue in experimental and theoretical particle physics:
comprehensive tests of a possible non-minimal fundamental scalar
sector, for which there exists a plethora of motivations. While there
has been a lot of progress in developing combined Higgs and gauge
analysis strategies for the LHC
Run~II~\cite{barca,legacy1,legacy2,other}, there exists no general and
proven analysis framework even for a Higgs portal model~\cite{portal}.

The announcement of an excess seen in the di-photon spectrum by both
ATLAS and CMS~\cite{750ex,S750_1000,S750_2100}, if confirmed by future
data, suggests such an extended scalar sector.  The anomaly has led to
an excessive number of publications, so we feel that adding one more,
and hopefully useful publication can be justified
somehow\footnote{Beyond the di-photon anomaly we present the first
  full \textsc{SFitter} analysis of a Higgs portal allowing for
  higher-dimensional operators.}.  Early studies of the anomaly in an
effective theory framework can be found in
Ref.~\cite{not_eft}. Intriguingly, an additional scalar is not
sufficient to explain the signal in complete models. The new scalar's
sizable couplings to photons and gluons need to be induced by
relatively light new particles~\cite{not_vectors}. For example in
supersymmetric models, vector-like matter added to the MSSM or
non-trivial signatures in the NMSSM are necessary for a successful
explanation of the excess~\cite{not_susy}. In models in which these
new states are connected to the SUSY breaking sector the new scalar
can be identified with the sgoldstino, implying a very low SUSY
breaking scale~\cite{not_sgoldstino}. Other extended spacetime
symmetries give rise to dilaton~\cite{not_dilaton} and radion
interpretations~\cite{not_radion}, which imply similarly unintended
consequences, such as low ultraviolet (UV) scales or a very large
curvature of the extra dimension. Extra dimensional scalars not
directly related to the compactification circumvent this
problem~\cite{not_extrad1} and can explain the localization of extra
dimensional fermions, which makes the new scalar a localizer
field~\cite{not_extrad2}. Related models, which consider the
electroweak scale (or the TeV scale) arising from composite dynamics
are less constrained than the MSSM, due to the large number of
potential scalar resonances and fermionic quark
partners~\cite{not_composite}. The possibility of the new resonance to
be a spin 2 particle, associated with a higher dimensional theory of
gravity is strongly constrained by dilepton searches and just like the
radion implies sizable curvature terms~\cite{not_spin2}. The large
width of $\Gamma_S=45$~GeV, as reported by ATLAS, can be addressed in
some models~\cite{not_interference1,not_wide}; while such a large
width only slightly increases the statistical significance, if it is
true, background interference effects are
important~\cite{not_interference2}. In this case, it is well motivated
to assume that the new scalar provides a portal to a dark sector,
inducing a sizable width through invisible decays~\cite{not_dm}.
Alternatively, it might be the sign of cascade decays or other
explanations not based on a single scalar resonance, which lead to
cusps and endpoint structures that can fake a large
width~\cite{not_exotics}. The very minimal, yet not perturbatively
realizable assumption of photon fusion induced production can not
explain a large width~\cite{not_photon}. The new resonance could also
be related to the various, persistent flavor
anomalies~\cite{not_flavor}, to the mechanism behind the electroweak
phase transition~\cite{not_phase}, the strong CP
problem~\cite{not_strong}, or an underlying string
theory~\cite{not_strings}. Finally, a variety of models, motivated by
different extensions of the Standard Model (SM) not fitting in the
above categories, and further measurements testing the properties of
the new resonance have been proposed~\cite{not_stuff}.\medskip

In spite of all these considerations, the most obvious question is
whether such an additional, likely scalar resonance can be part of an
extended Higgs sector~\cite{not_higgs,Dawson:2016ugw}; in other words,
if the new scalar can form a Higgs portal, possibly to a new
sector. To answer this question we will remain agnostic about the
underlying physics, but assume that a resonantly produced narrow
scalar singlet is responsible for the excess. We couple the new scalar
to the SM through an effective Lagrangian. This assumption exactly
corresponds to recent developments on how to describe deviations from
the Standard Model Higgs sector at the
LHC~\cite{eftfoundations,eftorig,higgsreview,barca,legacy1,legacy2,other,Grzadkowski:2010es}. The
combined Higgs portal Lagrangian is organized by the field content,
the symmetry structure, and the mass dimension.  This way we can
contrast the apparent absence of dimension--six effects in the range
$\Lambda \approx 300~...~500$~GeV for the SM-like Higgs
and gauge sector~\cite{legacy2} with the need for higher-dimensional operators
coupling to the new scalar with $\Lambda \lesssim 1$~TeV.

\subsection{Theoretical framework}
\label{sec:lag}

The most general linear effective Lagrangian up to dimension six and
built from Standard Model particles and a new scalar singlet reads
\begin{align}
\lag 
= \lag_\text{SM} + \lag_{\text{dim-6}}^H
+ \lag_{\text{dim}\leq 5}^S+\lag_{\text{dim-6}}^S \; .
\label{eq:lag1}
\end{align}
Here, $\lag_\text{SM}$ stands for the renormalizable SM Lagrangian,
while $\lag_{\text{dim-6}}^H$ contains the dimension--six operators made
out of SM fields. Adopting the basis of our set of Higgs legacy
papers~\cite{barca,legacy1,legacy2} it reads
\begin{align}
\lag_\text{dim-6}^H = &
  \frac{f_{BB}}{\Lambda^2} \phi^{\dagger} \hat{B}_{\mu \nu} \hat{B}^{\mu \nu} \phi
+ \frac{f_{WW}}{\Lambda^2} \phi^{\dagger} \hat{W}_{\mu \nu} \hat{W}^{\mu \nu} \phi 
- \frac{\alpha_s }{8 \pi} \frac{f_{GG}}{\Lambda^2} \phi^\dagger \phi G_{\mu\nu}^a G^{a\mu\nu}  
+ \frac{f_{WWW}}{\Lambda^2} \text{tr} \left( \hat{W}_{\mu \nu} \hat{W}^{\nu \rho}  \hat{W}_\rho{}^\mu \right) \notag \\
& + \frac{f_B}{\Lambda^2} (D_{\mu} \phi)^{\dagger}  \hat{B}^{\mu \nu}  (D_{\nu} \phi) 
+ \frac{f_W}{\Lambda^2} (D_{\mu} \phi)^{\dagger}  \hat{W}^{\mu \nu}  (D_{\nu} \phi) 
+ \frac{f_{\phi,2}}{\Lambda^2} \frac{1}{2} \partial^\mu ( \phi^\dagger \phi )
                            \partial_\mu ( \phi^\dagger \phi ) \notag \\
& + \left(\frac{f_\tau m_\tau}{v \Lambda^2} (\phi^\dagger\phi)(\bar L_3 \phi e_{R,3}) 
+ \frac{f_b m_b}{v \Lambda^2} (\phi^\dagger\phi)(\bar Q_3 \phi d_{R,3}) 
+ \frac{f_t m_t}{v \Lambda^2} (\phi^\dagger\phi)(\bar Q_3 \tilde \phi u_{R,3})+\text{h.c.}\right) \;  .
\label{eq:ourleff}
\end{align}
The Higgs covariant derivative is $D_\mu\phi= \left(\partial_\mu+ i g'
B_\mu/2 + i g \sigma_a W^a_\mu/2 \right)\phi $, and the field
strengths are $\hat{B}_{\mu \nu} = i g' B_{\mu \nu}/2$ and
$\hat{W}_{\mu\nu} = i g\sigma^a W^a_{\mu\nu}/2$ in terms of the Pauli
matrices $\sigma^a$. The $SU(2)_L$ and $U(1)_Y$ gauge couplings are
$g$ and $g^\prime$, respectively. While the minimum independent set consists of
59 baryon number conserving operators, barring flavor structure and Hermitian
conjugation~\cite{Grzadkowski:2010es}, we follow the definition of the
relevant operator basis describing Higgs coupling and triple gauge
boson vertex (TGV) modifications at the LHC in Ref.~\cite{barca}. In our construction we
assume a narrow, CP-even Higgs, focusing on the minimal, Yukawa-like, couplings to the heavy
fermions. We use the equations of motion to rotate
to a basis where there are no blind directions linked to electroweak precision data.
That way, we neglect all operators contributing to electroweak precision
observables at tree level in our LHC analysis.
For the Standard Model fit~\cite{barca,legacy1,legacy2} we
omit the operator $(\phi^\dagger \phi )^3$, which only contributes to
the rather poorly measured triple Higgs coupling. In the appendix we
argue why even in the presence of an additional, mixing scalar, this
operator will not add any extra relevant features to the fit.\medskip

Moving to the new scalar Lagrangian terms, we assume in the following
that the additional singlet does not develop a VEV, or that the
Lagrangian can be re-defined such that the VEV
vanishes~\cite{Dawson:2016ugw}. The effective Lagrangian of such an
additional singlet scalar can be divided into two pieces.  Following
Refs.~\cite{Gripaios:2016xuo,Kamenik:2016tuv,Bauer:2016lbe,Franceschini:2016gxv}
we first write down a set of non-redundant, independent operators up
to dimension five,
\begin{align}
\lag_{\text{dim}\leq 5}^S =&
 \frac{1}{2}\partial_\mu S \, \partial^\mu S-a_1 S-\frac{M_S^2}{2}S^2-a_3 S^3-a_4 S^4
-\frac{f_5^S}{\Lambda} S^5\notag\\
&-\mu_S S\phi^\dagger\phi
 -\frac{\lambda_{SH}}{2}S^2\phi^\dagger\phi 
 -\frac{f_1^S}{\Lambda}S ( \phi^\dagger\phi )^2
 -\frac{f_3^S}{\Lambda} S^3 \phi^\dagger\phi \notag\\
&+\frac{\alpha_s}{4\pi}\frac{f_{GG}^S}{\Lambda}SG_{\mu\nu}^aG^{a\;\mu\nu}
 +\frac{\alpha}{4\pi c_w^2}\frac{f_{BB}^S}{\Lambda}SB_{\mu\nu}B^{\mu\nu}
 +\frac{\alpha}{4\pi s_w^2}\frac{f_{WW}^S}{\Lambda}SW_{\mu\nu}^aW^{a\;\mu\nu} \notag \\
&\left(-\frac{f_d^S}{\Lambda}S\bar{Q}_L\phi d_R-\frac{f_u^S}{\Lambda}S\bar{Q}_L\tilde\phi u_R
 -\frac{f_\ell^S}{\Lambda}S\bar{L}_L\phi \ell_R+\text{h.c.}\right)\;\;\;.
\label{eq:lagd5}
\end{align}
%

To be fully consistent with the Standard Model Lagrangian we could
then add all dimension--six operators including at least one power of the
new singlet scalar. The corresponding set of additional
operators can be written as~\cite{Gripaios:2016xuo}
\begin{align}
\lag_{\text{dim-6}}^S =& 
 \frac{f^{SS}_\phi}{\Lambda^2} \phi^\dagger\phi \partial_\mu S \partial^\mu S
-\frac{f_6^S}{\Lambda^2} S^6
-\frac{f_4^S}{\Lambda^2} S^4 \phi^\dagger\phi
-\frac{f_2^S}{\Lambda^2} S^2 (\phi^\dagger\phi )^2 \notag \\
&+\frac{f_{GG}^{SS}}{\Lambda^2}S^2 G_{\mu\nu}^aG^{a\;\mu\nu}
+\frac{f_{BB}^{SS}}{\Lambda^2}S^2B_{\mu\nu}B^{\mu\nu}
+\frac{f_{WW}^{SS}}{\Lambda^2}S^2W_{\mu\nu}^aW^{a\;\mu\nu} \notag \\
&\left(-\frac{f_d^{SS}}{\Lambda^2}S^2\bar{Q}_L\phi d_R-\frac{f_u^{SS}}{\Lambda^2}S^2\bar{Q}_L\tilde\phi u_R
-\frac{f_\ell^{SS}}{\Lambda^2}S^2\bar{L}_L\phi \ell_R+\text{h.c.}\right)\;\;.
\label{eq:lagd6}
\end{align}
%
Nevertheless, given the singlet
nature of the new scalar and neglecting lepton number violation, all
dimension--six operators including the singlet are quadratic in the field
$S$. Consequently, their phenomenological effects will be
contributions to the mass terms ($f_2^S/\Lambda^2$), re-definitions of
the $S$ field to recover canonical kinetic terms
($f^{SS}_\phi/\Lambda^2$), and the contributions to several vertices
including two or more heavy scalars. After scalar-Higgs mixing, the
two operators $f^{SS}_\phi/\Lambda^2$ and $f^S_2/\Lambda^2$ will contribute
to the $SHH$ interaction as well. However, all these
phenomenology features are already taken into account in our analysis
by the free parameters in the dimension--five Lagrangian. Therefore, we
neglect for the time being the explicit features induced by
Eq.\eqref{eq:lagd6}. We give more details on the effective Lagrangian
and the Higgs portal mixing in the Appendix.

\subsection{Analysis framework}
\label{sec:sfitter}

The set of analyses presented here are derived using the
\textsc{SFitter} framework. \textsc{SFitter} allows us to study
multi-dimensional parameter spaces in the Higgs
sector~\cite{sfitter_higgs,legacy1}, the gauge sector~\cite{legacy2}
and in new physics models like supersymmetry~\cite{sfitter_susy}. The
fit procedure uses Markov chains to create an exclusive,
multidimensional log-likelihood map, based on the available measurements
and including all the relevant uncertainties and correlations. The
construction of a profile likelihood with flat theory uncertainties
leads to the \textsc{RFit} scheme~\cite{rfit}. The statistic uncertainties on
the measurements, both for event rates and kinematic distributions,
follow Poisson statistics, as do the background uncertainties. All
systematic uncertainties are described by Gaussian distributions and
can be correlated between the relevant channels. We show
log-likelihood projections on two-dimensional planes after profiling
over all other parameters. Here, red-yellow regions will illustrate
points within $\Delta (-2\log \mathcal{L})=2.3$ of the best fit point
log-likelihood ($1\sigma$ in the Gaussian approximation), green
regions indicate $\Delta (-2\log {\cal L})=6.18$ ($2\sigma$ in the
Gaussian limit), and black dots imply the $\Delta (-2\log
\mathcal{L})=5.99$ exclusion limits (95\%~CL in the Gaussian
case).\medskip

\begin{table}[b!]
\begin{tabular}{llc}
\hline
Channel &  Dataset & Reference \\
\hline
$ S \rightarrow \gamma \gamma$   & ATLAS  8 TeV & \cite{S750_1000} \\
$ S \rightarrow \gamma \gamma$   & ATLAS 13 TeV & \cite{S750_1000} \\
$ S \rightarrow \gamma \gamma$   & CMS    8 TeV & \cite{S750_2000} \\
$ S \rightarrow \gamma \gamma$   & CMS   13 TeV & \cite{S750_2100} \\
$ S \rightarrow W W$             & ATLAS  8 TeV & \cite{S750_1010} \\
$ S \rightarrow W W$             & ATLAS 13 TeV & \cite{S750_1110} \\
$ S \rightarrow Z Z$             & ATLAS  8 TeV & \cite{S750_1020} \\
$ S \rightarrow Z Z$             & ATLAS 13 TeV & \cite{S750_1120} \\
$ S \rightarrow Z Z$             & ATLAS 13 TeV & \cite{S750_1121} \\
$ S \rightarrow Z \gamma$        & ATLAS 13 TeV & \cite{S750_1130} \\
$ S \rightarrow Z \gamma$        & CMS   13 TeV & \cite{S750_2130} \\
$ S \rightarrow Z \gamma$        & ATLAS  8 TeV & \cite{S750_1030} \\
$ S \rightarrow t\bar{t}$        & ATLAS  8 TeV & \cite{S750_1040} \\
$ S \rightarrow j j$             & CMS    8 TeV & \cite{S750_2050} \\
$ S \rightarrow h h$             & ATLAS 13 TeV & \cite{S750_1160} \\
$ S \rightarrow h h$             & CMS    8 TeV & \cite{S750_2060} \\
$ S \rightarrow \tau \bar{\tau}$ & CMS    8 TeV & \cite{S750_2070} \\
\hline
\end{tabular}
\caption{Experimental data on the heavy resonance included in our
  fit.}
\label{tab:750_data}
\end{table}

The implementation of experimental results in the \textsc{SFitter}
framework is described in Ref.~\cite{legacy1} for the Higgs measurements 
and in Ref.~\cite{legacy2} for anomalous triple gauge boson coupling measurements.
For the triple gauge boson vertex (TGV) analyses\footnote{Note that pair production of weak bosons at the LHC
  is a crucial ingredient to a Higgs fit based on an effective
  Lagrangian assuming a linear realization of electroweak symmetry
  breaking. Without taking these measurements into account the
  qualitative and quantitative outcome of the fit will be
  wrong~\cite{legacy2}.}  the correlation of the theory uncertainties
between the different bins of a given kinematic distribution is taken
into account by flat profiled nuisance parameters~\cite{legacy2}, while for the
different Higgs channels the theory uncertainties are considered
uncorrelated without a sizable impact on the shown
results~\cite{legacy1}.  For the Higgs portal analysis we take into
account the constraints on a possible new resonance based on the data
listed in Tab.~\ref{tab:750_data}. For the new resonance we only
implement inclusive measurements assuming a narrow width.

\section{Higgs portal analysis}
\label{sec:ana}

In the following we will use the \textsc{SFitter} effective Lagrangian
framework to analyze a new gluon-fusion produced resonance in
combination with the electroweak gauge and Higgs sectors at the weak
scale. In other words, we ask the question whether such a new particle
could be part of an extended Higgs sector and what the allowed
parameter space is. In Sec.~\ref{sec:ana_heavy} we only include the
dimension--five operators given in Eq.\eqref{eq:lagd5}, restricting
the analysis to the data in Tab.~\ref{tab:750_data}. In
Sec.~\ref{sec:ana_combined} we combine this analysis with the
Higgs-electroweak measurements and the \textsc{SFitter} results
induced by the dimension--six Lagrangian in
Eq.\eqref{eq:ourleff}. Finally we link the size of different operators
to a common origin in Sec.~\ref{sec:ana_magic}.

\subsection{Heavy scalar fit}
\label{sec:ana_heavy}

\begin{figure}[t]
\includegraphics[width=0.32\textwidth]{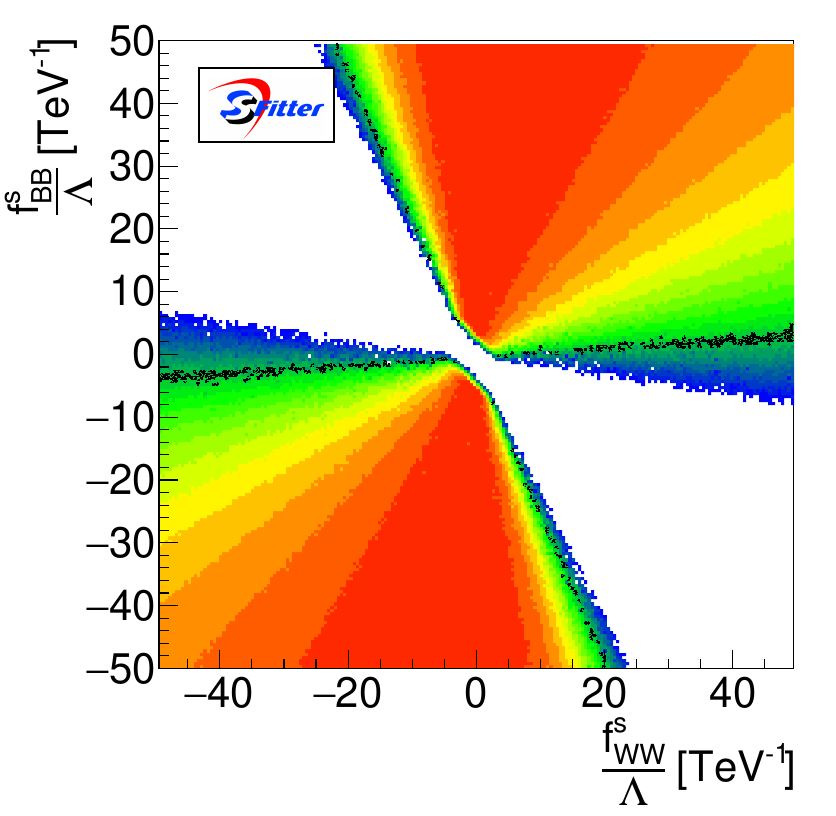}
\hspace{-0.3cm}
\includegraphics[width=0.32\textwidth]{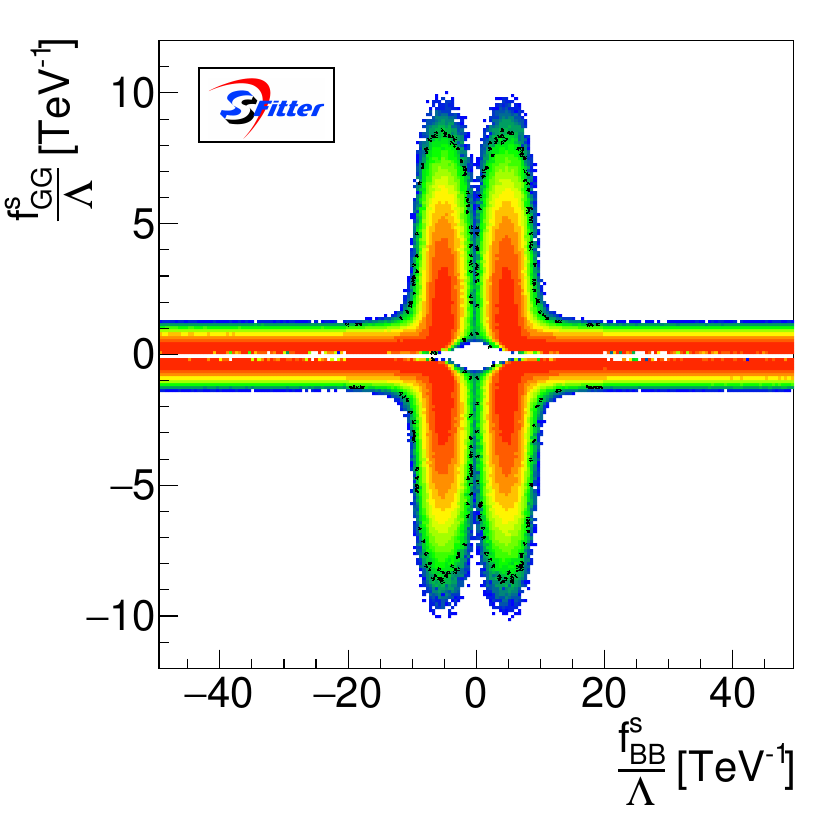}
\hspace{-0.3cm}
\includegraphics[width=0.32\textwidth]{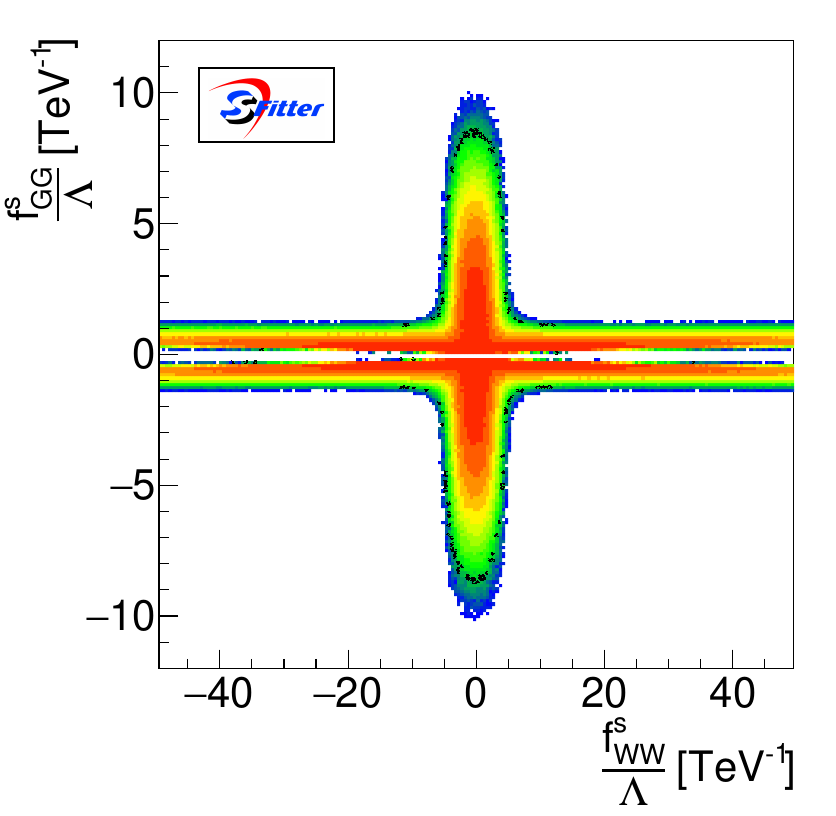}
\hspace{-0.2cm}\raisebox{20pt}{\includegraphics[width=0.047\textwidth]{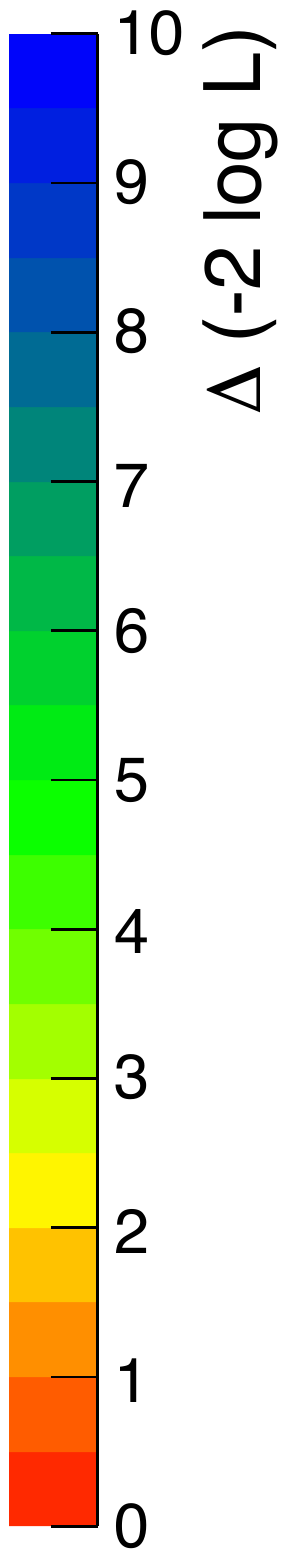}}\\
\includegraphics[width=0.32\textwidth]{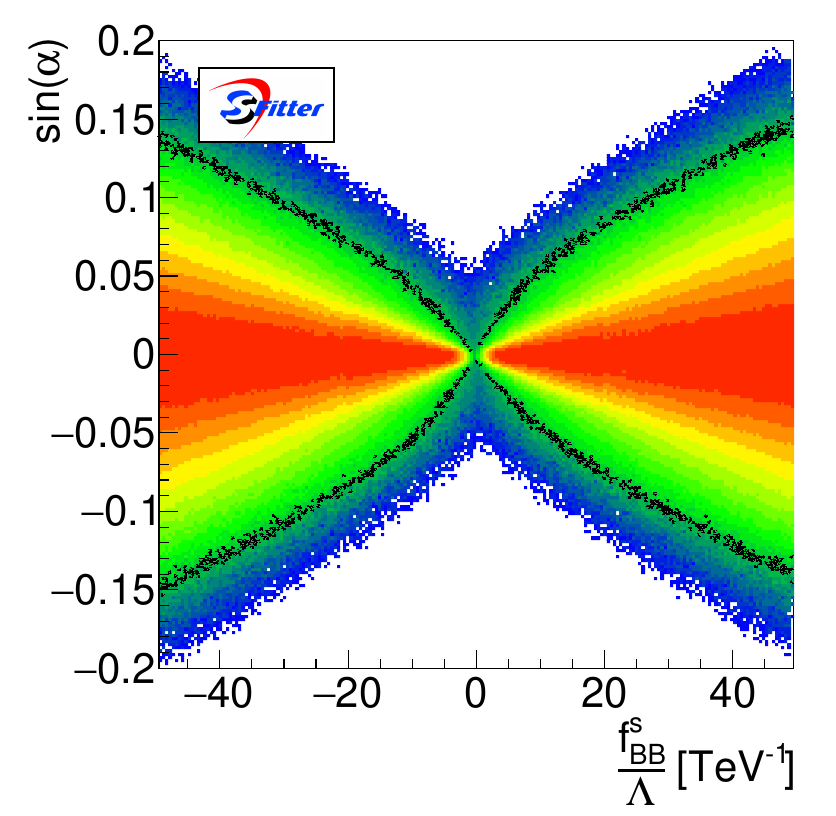}
\hspace{-0.3cm}
\includegraphics[width=0.32\textwidth]{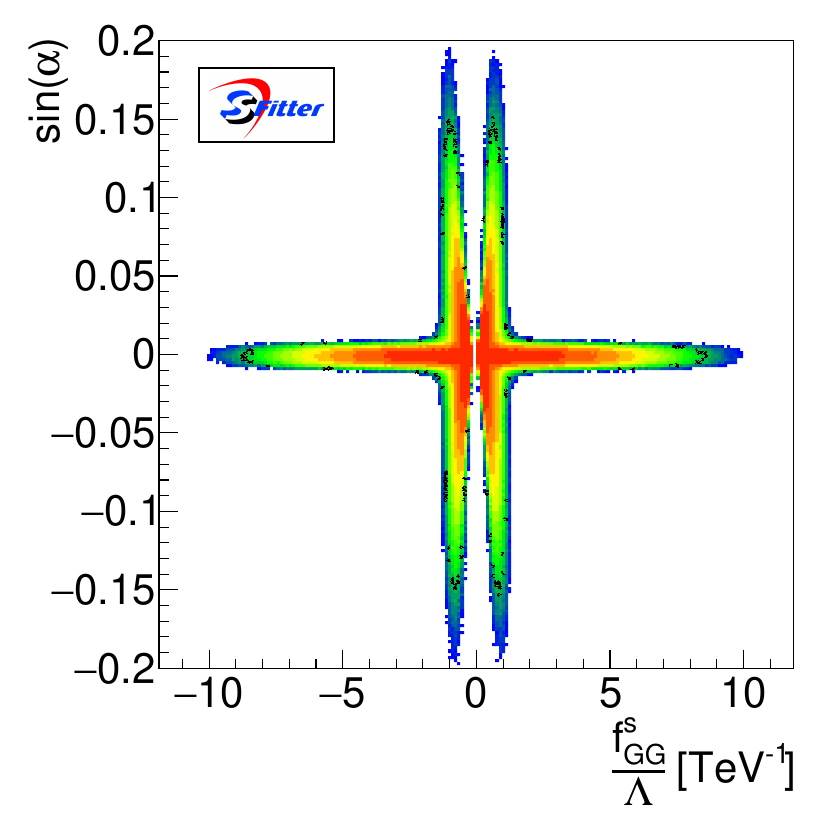}
\hspace{-0.3cm}
\includegraphics[width=0.32\textwidth]{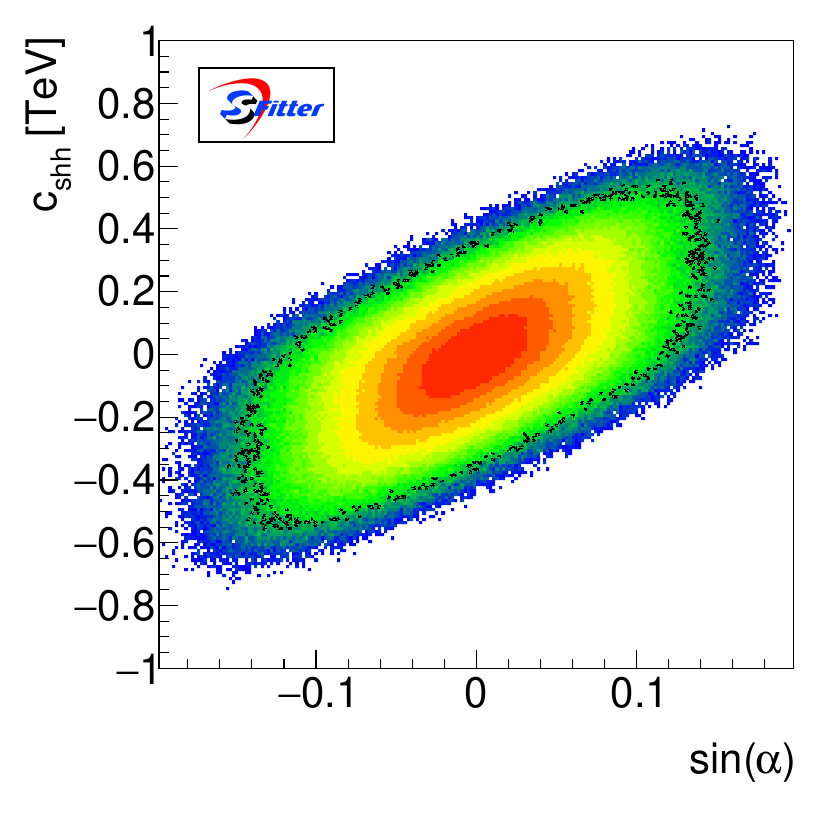}
\hspace{-0.2cm}\raisebox{20pt}{\includegraphics[width=0.047\textwidth]{Figures/barcode.pdf}}
\caption{Two-dimensional profile log-likelihoods for the analysis of
  the heavy scalar sector alone spanning $f^S_{WW}$, $f^S_{BB}$,
  $f^S_{GG}$, $\sin\alpha$, and $c_{SHH}$. The black points indicate
  $\Delta(-2\log L)=5.99$.}
\label{fig:5param_750}
\end{figure}

As a first step we analyze only the measurements for the heavy scalar,
as listed in Tab.~\ref{tab:750_data}. In Fig.~\ref{fig:5param_750} we
use this data to determine the five parameters
\begin{align}
 \{ \, f^S_{WW}, f^S_{BB}, f^S_{GG}, \sin\alpha, c_{SHH} \, \} \; .
\end{align}
In our parametrization $c_{SHH}$ accounts for the independent
contributions to the $SHH$ vertex from the dimension--five Lagrangian
terms beyond the terms generating the mixing, as discussed in the
Appendix.

The best fit point for this analysis has $-2\log L=8.9$, while the SM
point leads to $-2\log L=28.2$, within a $3.1\sigma$ range for a 5-parameter
study (in the Gaussian limit). In the upper left panel of Fig.~\ref{fig:5param_750} we can see
that within the displayed range of parameters both $f^S_{WW}$ and
$f^S_{BB}$ are strongly correlated, and they present a flat direction.
The correlation reflects the fact that they are the
only Wilson coefficients contributing to the di-photon decay of the
new scalar at tree level. Because $f^S_{WW}$ is constrained through
the decay $S \rightarrow WW$, the di-photon excess cannot be
accommodated through this coupling only. Due to that we find 
that $|f^S_{BB}/\Lambda| > 2~\itev$ in the upper-center panel. 
This is caused by the fact that $f^S_{BB}$ does not contribute to the $SWW$ vertex, and in
addition its contribution to the $SZZ$ vertex is suppressed by the weak mixing
angle. This allows us to explain the observed excess without getting
into conflict with the exclusion bounds, what makes $f^S_{WW}=0$
compatible with the best fit point, as shown in the upper-right panel.

Moving on to the mixing angle, we find $\sin \alpha < 0.15$ at 95\%~CL
and for the displayed ranges of $f^S_{BB}$ and $f^S_{WW}$ in the
lower-left panel. This bound comes from the absence of a heavy scalar
signal in $WW$ and $ZZ$, but also in di-jet, $t\bar{t}$, $\tau\bar{\tau}$, and $hh$
decay channels. It is linked to maximum assumed values for $f^S_{BB}$
and $f^S_{WW}$, because a larger mixing angle can be partially compensated by
larger Wilson coefficients $f^S_{BB}$+$f^S_{WW}$. For large values the
di-photon branching ratio of the heavy scalar can exceed 50\%, while the
remaining decay channel modes are suppressed, allowing $\sin \alpha$ to increase
without conflicting with data. If
we allow for extreme values of $f^S_{BB}/\Lambda+f^S_{WW}/\Lambda \sim
250~\itev$, the upper bound on $\sin \alpha$ goes up to $0.3$. In the
lower-center panel we again observe two distinct regions in
$f^S_{GG}$. The vertical region with $f^S_{GG}/\Lambda < 1.5$ TeV$^{-1}$ is characterized
by a large branching ratio for $S\rightarrow \gamma \gamma$, linked to
large values of $f^S_{BB}+f^S_{WW}$.  The horizontal region with
$f^S_{GG}/\Lambda = 1.5~...~10$ TeV$^{-1}$ is characterized by a large production rate
for the new scalar and a total decay width driven by $f^S_{GG}$.
The upper limit on $f^S_{GG}$ is set by di-jet
searches, and the mixing in this regime has to be small to respect the
limits from other decay channels. Finally, in the lower-right panel
we show the correlation between the mixing angle and $c_{SHH}$ from
the limit on the decay $S\rightarrow HH$. Fixing $c_{SHH} = 0$ and
generating the $SHH$ interaction through the mixing angle alone has no
effect on any of the other correlations.\medskip

\begin{figure}[t]
\includegraphics[width=0.32\textwidth]{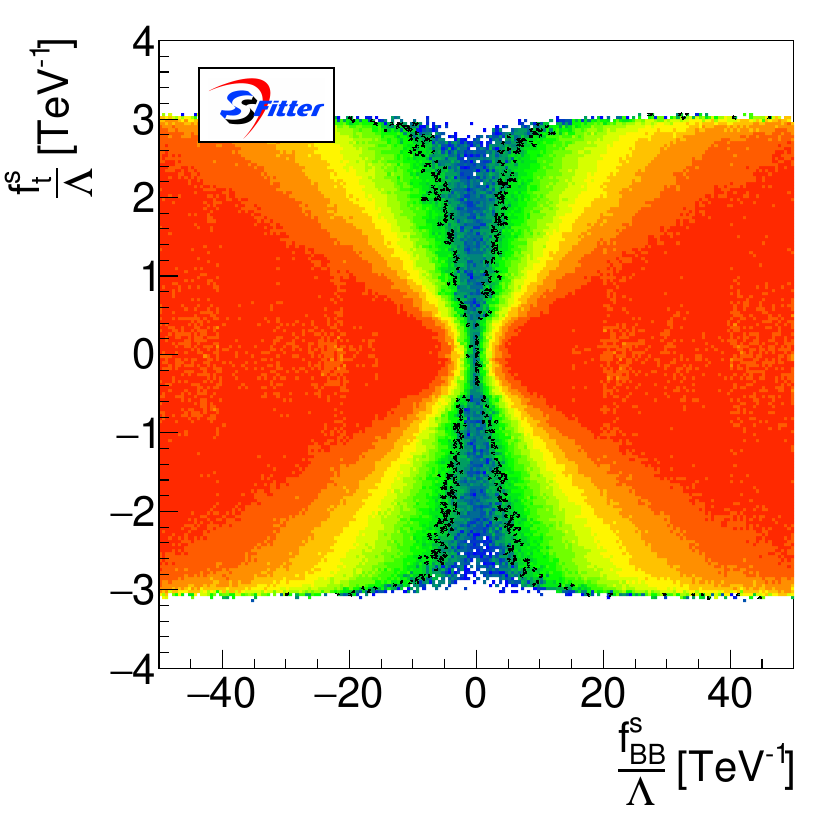}
\hspace{-0.3cm}
\includegraphics[width=0.32\textwidth]{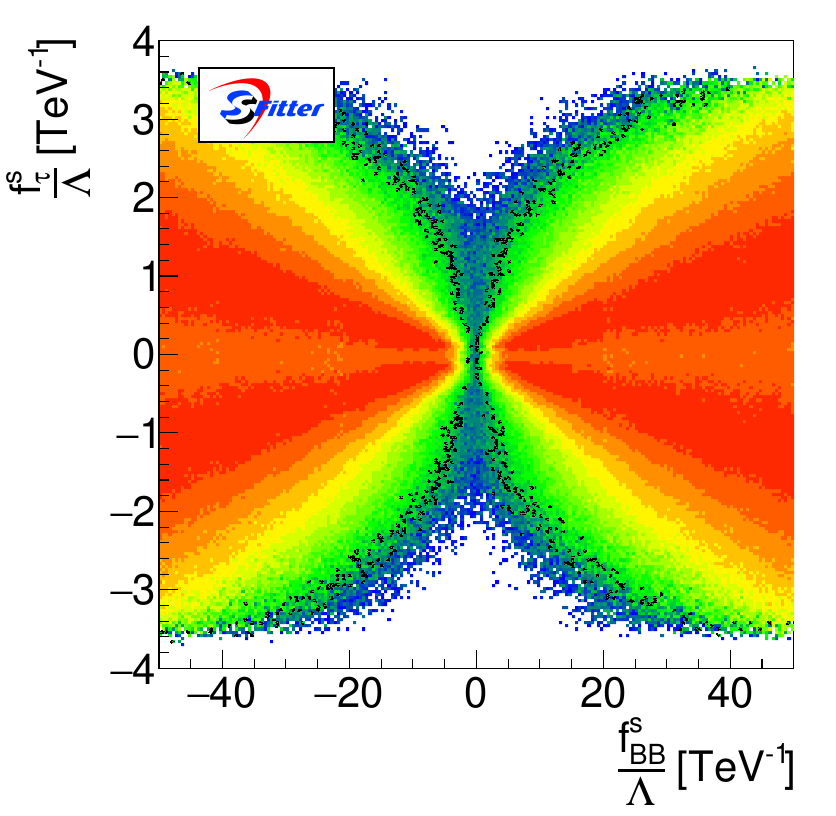}
\hspace{-0.3cm}
\includegraphics[width=0.32\textwidth]{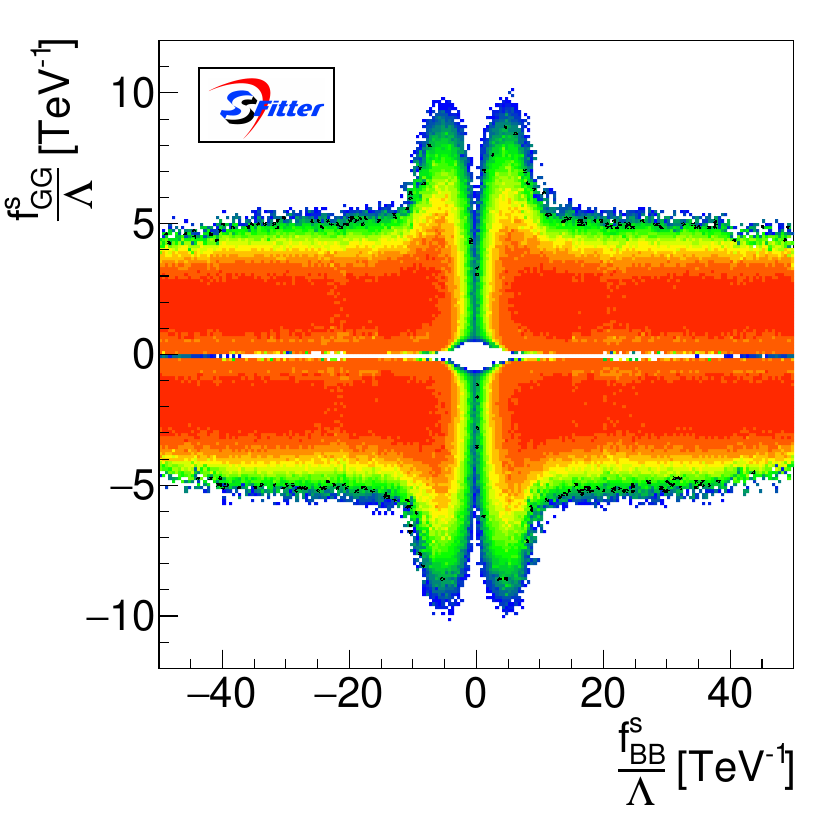}
\hspace{-0.2cm}\raisebox{20pt}{\includegraphics[width=0.047\textwidth]{Figures/barcode.pdf}}\\
\includegraphics[width=0.32\textwidth]{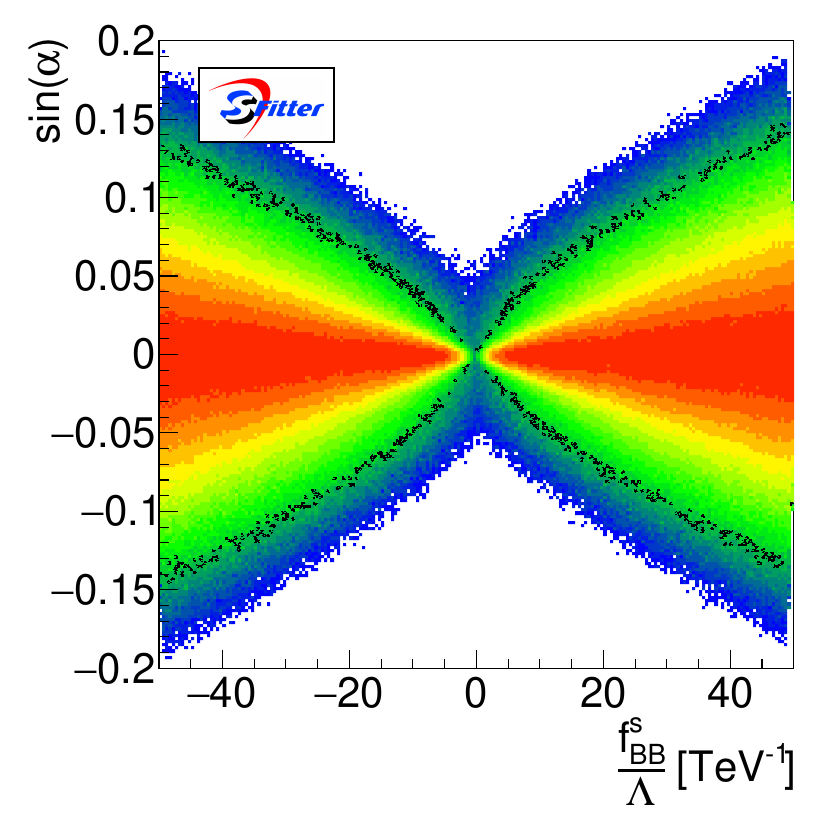}
\hspace{-0.3cm}
\includegraphics[width=0.32\textwidth]{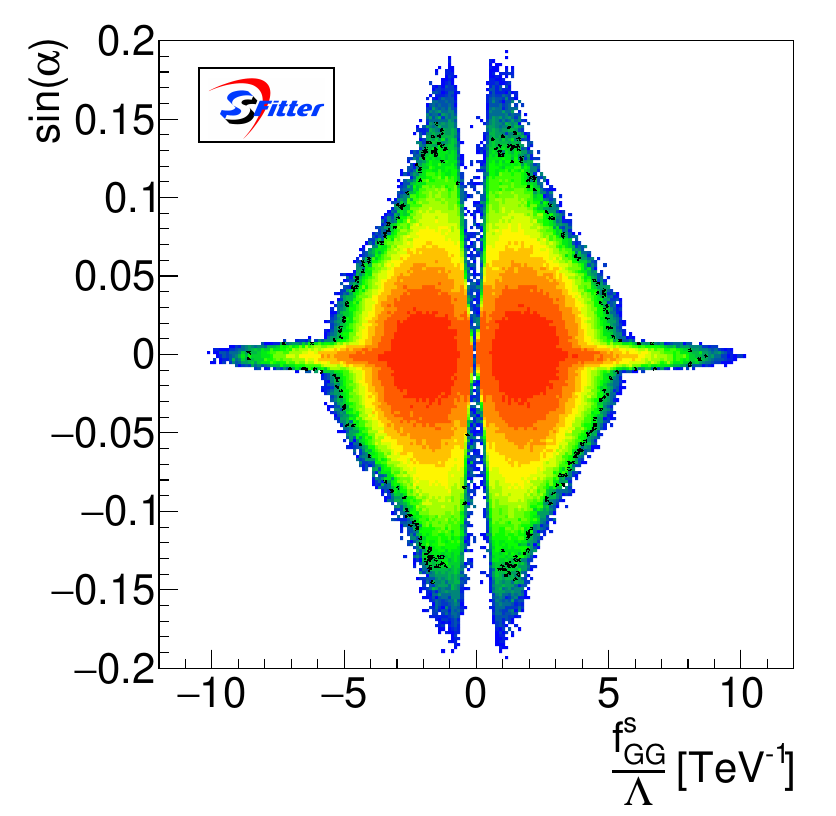}
\hspace{-0.3cm}
\includegraphics[width=0.32\textwidth]{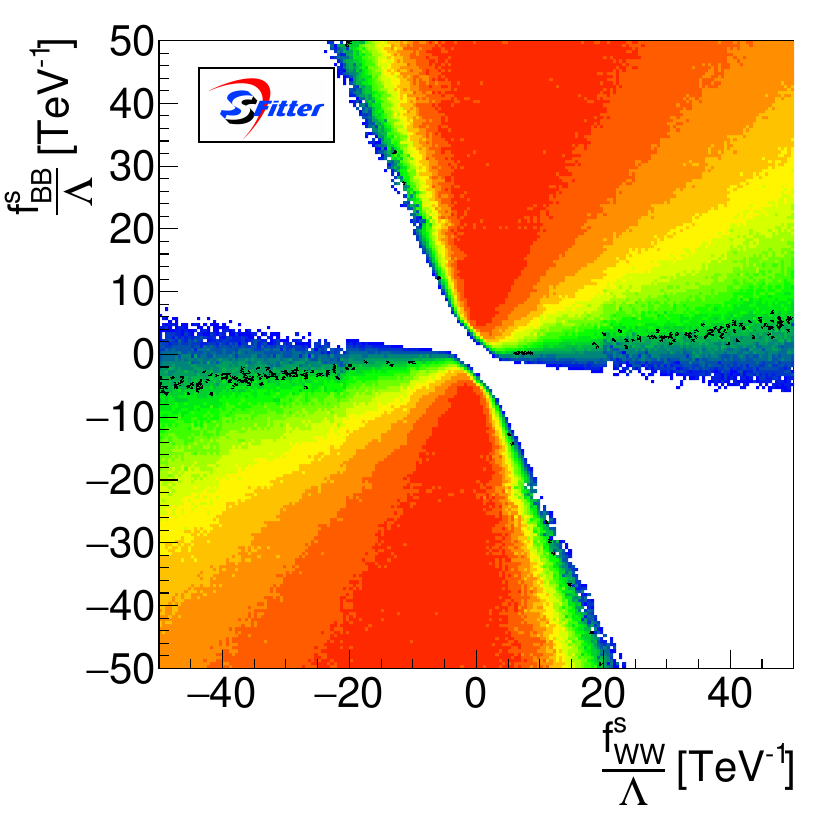}
\hspace{-0.2cm}\raisebox{20pt}{\includegraphics[width=0.047\textwidth]{Figures/barcode.pdf}}\\
\caption{Two-dimensional profile log-likelihoods for the analysis of
  the heavy scalar sector alone.  In contrast to
  Fig.~\ref{fig:5param_750} we now include fermion couplings in our set
  of Wilson coefficients $f^S_{WW}$, $f^S_{BB}$, $f^S_{GG}$,
  $\sin\alpha$, $c_{SHH}$, $f^S_{t}$ and $f^S_{\tau}$.  The black
  points indicate $\Delta(-2\log L)=5.99$.}
\label{fig:7param_750}
\end{figure}

We proceed with an analysis allowing the new scalar to couple
to the two fermions for which there are direct searches available. The
analysis now includes
\begin{align}
 \{ \, f^S_{WW}, f^S_{BB}, f^S_{GG}, \sin\alpha, c_{SHH}, f^S_{t}, f^S_{\tau} \, \} \; .
\end{align}
A selection of results is shown in Fig.~\ref{fig:7param_750}. The
fermionic Wilson coefficients $f_t^S$ and $f_\tau^S$ are constrained
by $t\bar{t}$ and $\tau^+\tau^-$ resonance searches, as well as from
an upper limit $\Gamma_S < 25$~GeV which we assume throughout our
analysis and which sets hard limits on $f_t^S$ and $f_\tau^S$. The
best fit point of this run is only mildly better than before, $-2\log
L=8.3$. The limits on these two fermion couplings are stronger for
smaller $f_{BB}^S$, as illustrated for $f_t^S$ in the upper-left
panel, and $f_{\tau}^S$ in the upper-center one. The reason is that in
those regions the partial decay width to photons becomes small, and
the required di-photon branching ratio translates into small fermionic
couplings.  Conversely, larger fermionic Wilson coefficients now allow
for best fit regions with large $f^S_{GG}$ and $f^S_{BB}$ at the same
time, as shown in the upper-right panel of
Fig.~\ref{fig:7param_750}. This is the main difference with respect to
the reduced analysis shown in Fig.~\ref{fig:5param_750}.  The rest of
correlations remain qualitatively unchanged, as can be seen in the
lower panels of Fig.~\ref{fig:7param_750}.  In particular the upper
95\% CL limit on the mixing angle is still $\sin \alpha < 0.15$.\medskip

In passing we note that all the results shown so far have been derived
assuming a CP-even new scalar. Nevertheless, for the analysis up to
this point the results remain unchanged when instead we assume a heavy
CP-odd scalar.

\subsection{Combined Higgs portal fit}
\label{sec:ana_combined}

\begin{figure}[t]
\includegraphics[width=0.32\textwidth]{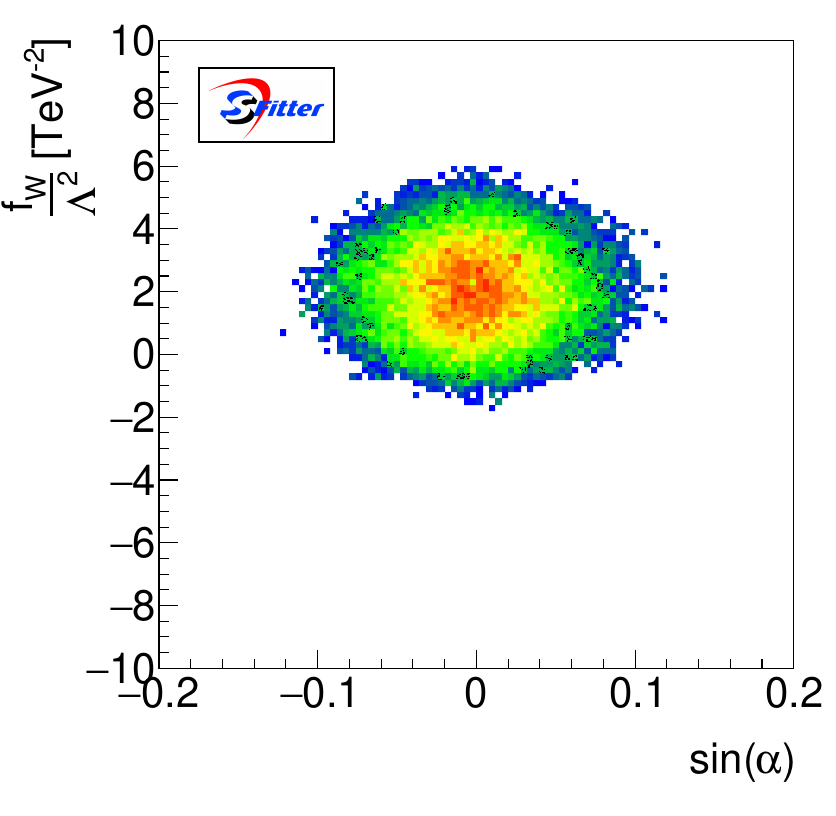}
\hspace{-0.3cm}
\includegraphics[width=0.32\textwidth]{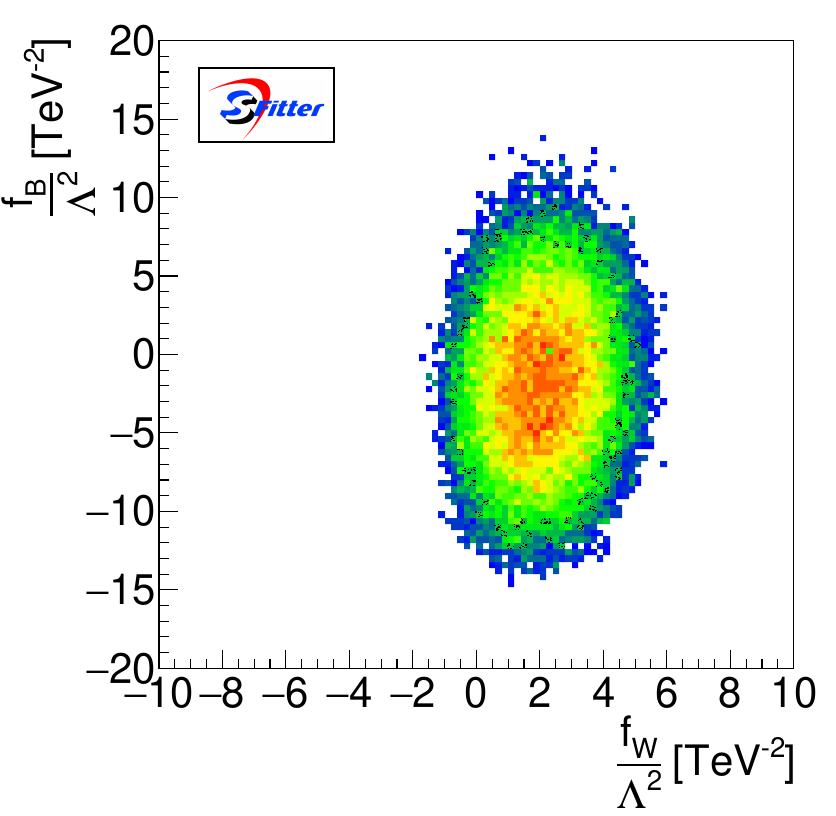}
\hspace{-0.3cm}
\includegraphics[width=0.32\textwidth]{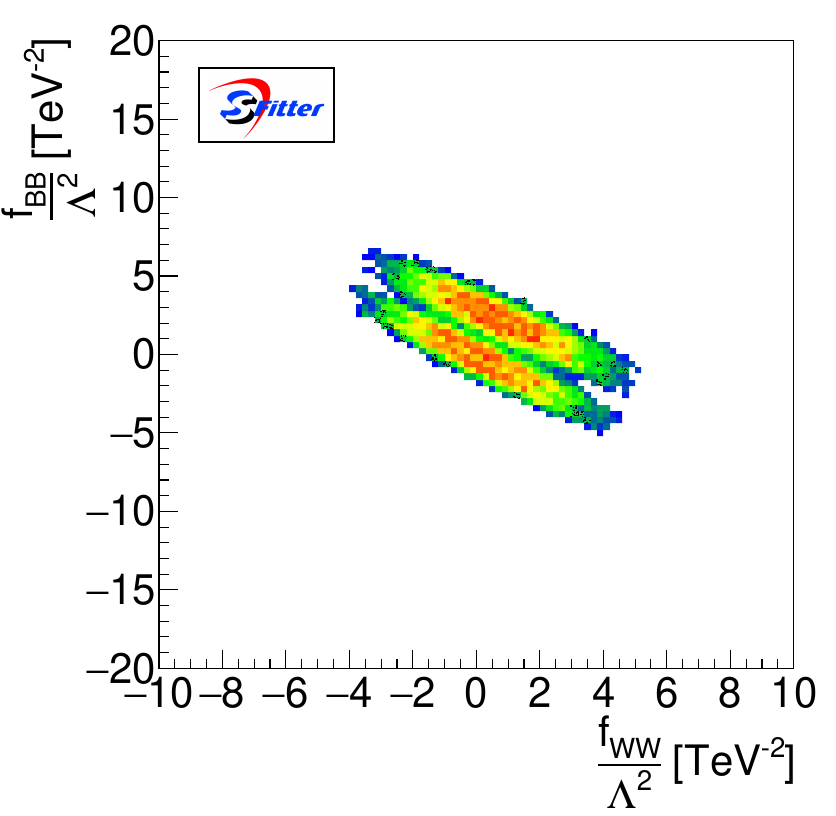}
\hspace{-0.2cm}\raisebox{20pt}{\includegraphics[width=0.047\textwidth]{Figures/barcode.pdf}}
\caption{Two-dimensional profile log-likelihoods for the combined
  Higgs, TGV, and heavy scalar sectors. The black points indicate
  $\Delta(-2\log L)=5.99$.}
\label{fig:17param_750}
\end{figure}

Next, we discuss the results for the general scenario, where we 
constrain the 17 parameters
\begin{align}
\{ \, f^S_{WW}, f^S_{BB}, f^S_{GG}, \sin\alpha, f^S_t, f^S_b, f^S_\tau,
  f_{WW}, f_{BB}, f_{GG}, f_{W}, f_B, f_{\phi,2}, f_{WWW}, f_t, f_b, f_\tau \, \}
\end{align}
from the combined measurements in the electroweak-Higgs, and
the heavy scalar sector. We have fixed $c_{SHH}=0$ given its minor
impact on the fit results.

In this case the best fit value has a likelihood of $-2\log L=242.0$,
for an analysis containing $252$ measurements, while the Standard
Model point leads to $-2\log L=273.9$.  In Fig.~\ref{fig:17param_750}
we show a reduced selection of correlations between Wilson
coefficients. When adding the heavy scalar to the combined Higgs and
gauge boson analysis, the potentially largest change in the results
appears for $f_W$ and $f_B$. The twofold reason is illustrated in
detail in the Appendix. First, focusing on the electroweak-Higgs
phenomenology, while the contribution of $f_W$ and $f_B$ to the Higgs
vertices is now weighted by the cosine of the mixing angle, their
contribution to the triple gauge boson vertex is not. This generates a
different pattern of Higgs-TGV correlations once we add the new
scalar. Second, the mixing of the Higgs boson with the heavy scalar
allows $f_W$ and $f_B$ to generate genuinely new Lorentz structure
contributions to the $SWW$, $SZZ$ and $SZ\gamma$ vertices, on top of
the contributions from the rest of dimension--five and dimension--six
operators.

The first effect turns out to be negligible, and given the small
allowed size for the mixing angle, the electroweak-Higgs measurements
are not precise enough to be sensitive to the scalar mixing
contributions.  Conversely, the second effect is more important.  The
mild preference for non-zero $f_W$ values from the electroweak-Higgs
measurements~\cite{legacy2} causes the best fit regions to generate
the new contribution to the decays $S \to WW, ZZ, Z\gamma$. These
channels can be then better fit suppressing them further with a
smaller mixing angle. The addition of the dimension--six operator
causes then the upper bound on the scalar mixing to be mildly reduced
with respect to the results in the previous section: now $\sin \alpha
< 0.10$ at 95\% CL.  This can be observed in the left panel of
Fig.~\ref{fig:17param_750}.

Apart from this effect, the small mixing angle causes a lack of sizable
correlations between both the new scalar sector and the electroweak-Higgs sector.
Consequently, the results and two-dimensional planes involving dimension--five
operators are very similar to the ones shown in
Fig.~\ref{fig:7param_750}. The planes involving dimension--six
operators remain unchanged with respect to the results shown in
Ref.~\cite{legacy2}, something that we illustrate in the center and
right panels of Fig.~\ref{fig:17param_750} for two of the
dimension--six correlations.

\subsection{A common origin of operators}
\label{sec:ana_magic}

When we split a common scalar potential for two mixing states into a
set of dimension--five and dimension--six operators, the question becomes how different the
higher-dimensional effects in the light and heavy scalar couplings can
really be. In this section we assume that the set of heavy scalar
couplings are directly tied to their Higgs-like counter parts,
\begin{alignat}{9}
\frac{f_{GG}}{\Lambda^2} &= -2 \frac{f_{GG}^S}{\Lambda} \left|\frac{f_{GG}^S}{\Lambda}\right|
\qqqquad
& \frac{f_{f}}{\Lambda^2} &=-\frac{v}{m_f}\frac{f_{f}^S}{\Lambda}\left|\frac{f_{f}^S}{\Lambda}\right| \notag\\ 
 \frac{f_{BB}}{\Lambda^2} & = -\frac{1}{4\pi^2}\frac{f_{BB}^S}{\Lambda}\left|\frac{f_{BB}^S}{\Lambda}\right|
& \frac{f_{WW}}{\Lambda^2} & = -\frac{1}{4\pi^2}\frac{f_{WW}^S}{\Lambda}\left|\frac{f_{WW}^S}{\Lambda}\right| \; ,
\label{eq:prior1}
\end{alignat}
for $f = b,t,\tau$.  The relative signs and pre-factors ensure that
the underlying new physics scales are consistent, as 
defined in Eq.\eqref{eq:lagd5}.
For the fermion case, this is motivated by the need to
have a minimal flavor violating structure in both dimension--five and dimension--six
operators to avoid large flavor changing neutral currents~\cite{Goertz:2014qia}.
In a Bayesian language this approach would correspond to a Dirichlet prior, for
example employed in the dark matter fit of Ref.~\cite{tim}, with an exponent
parameter $\alpha\gg1$.\medskip

After imposing the relations in Eq.\eqref{eq:prior1},
we proceed to perform the combined Higgs, triple gauge boson
vertex and heavy scalar analysis spanning the 11 free
parameters
\begin{align}
 \{ \, f^S_{WW}, f^S_{BB}, f^S_{GG}, f^S_t, f^S_b, f^S_\tau, \sin\alpha,
       f_{W}, f_B, f_{\phi,2}, f_{WWW} \, \} \; . 
\end{align}
We have again fixed $c_{SHH}=0$, while $f_{WW}$, $f_{BB}$, $f_{GG}$,
$f_t$, $f_b$ and $f_\tau$ are set from
Eq.\eqref{eq:prior1}. Interestingly, the best fit point is $-2\log
L=242.6$, \ie within the analysis precision very close to the best fit
point of the previous general scenario. This illustrates one of the
most important conclusions: when dimension--five and dimension--six
operators of a similar type are imposed to be related, there are still
regions in the new physics parameter space which can accommodate the 
di-photon anomaly while respecting the constraints from the
electroweak-Higgs measurements.\medskip

\begin{figure}[t]
\includegraphics[width=0.32\textwidth]{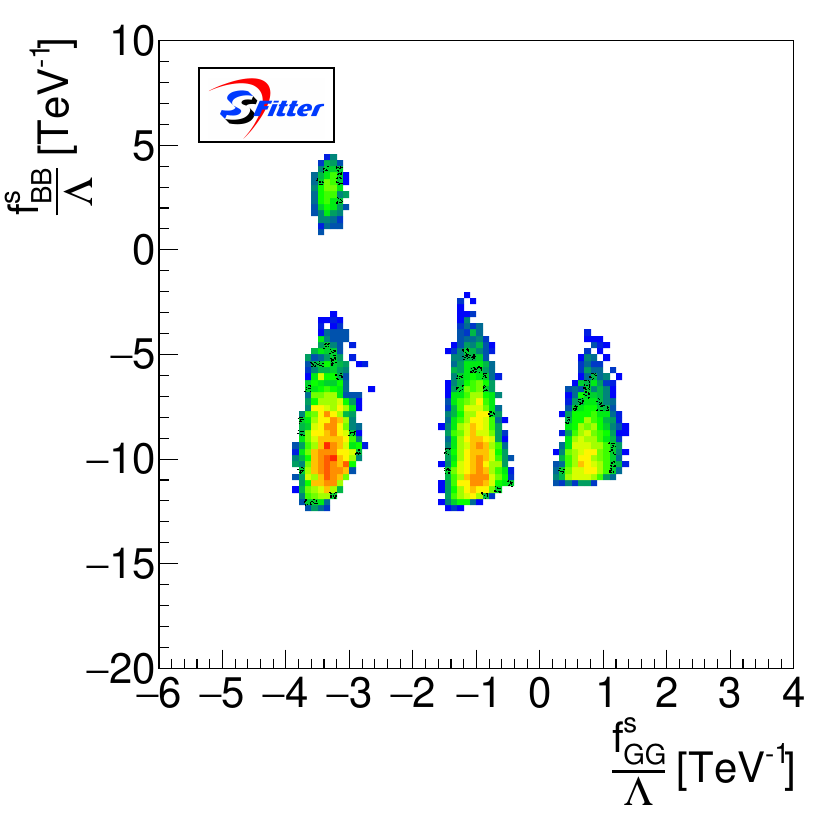}
\hspace{-0.3cm}
\includegraphics[width=0.32\textwidth]{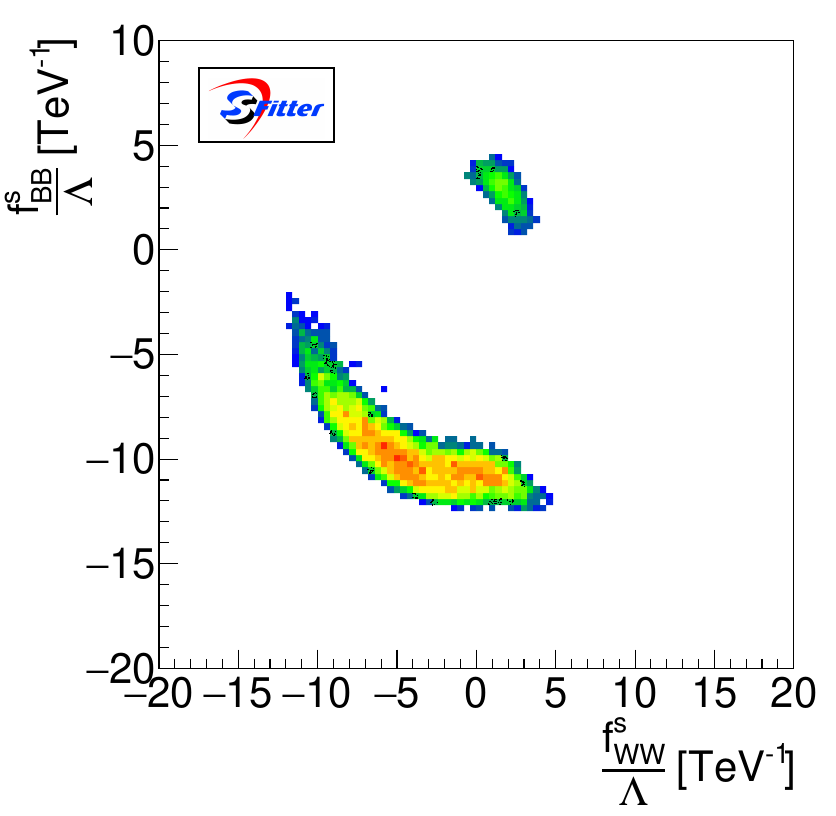}
\hspace{-0.3cm}
\includegraphics[width=0.32\textwidth]{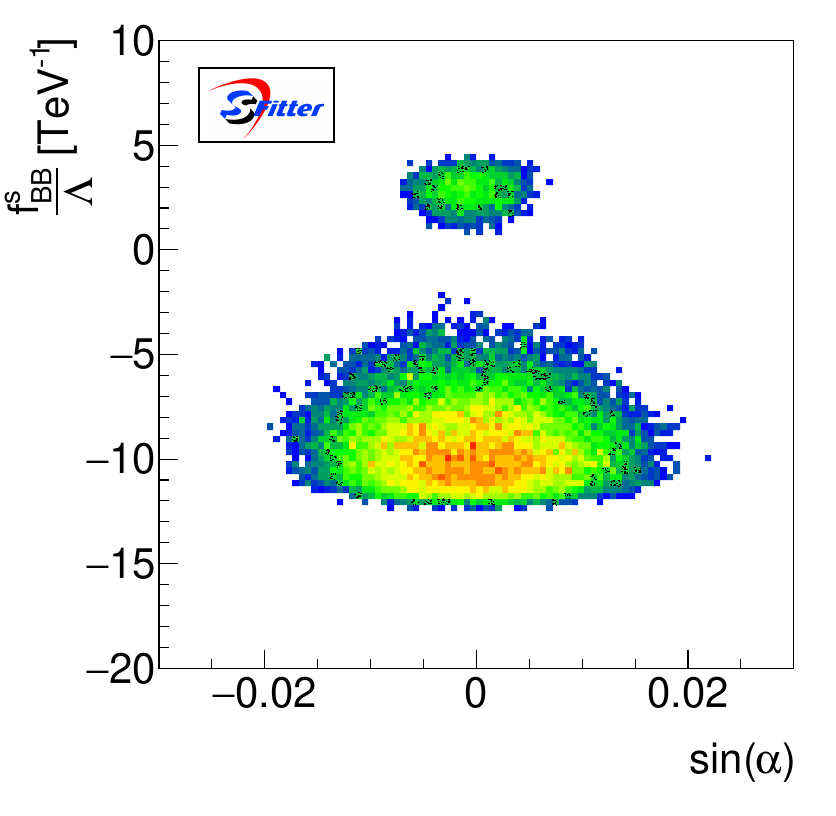}
\hspace{-0.2cm}\raisebox{20pt}{\includegraphics[width=0.047\textwidth]{Figures/barcode.pdf}}\\
\includegraphics[width=0.32\textwidth]{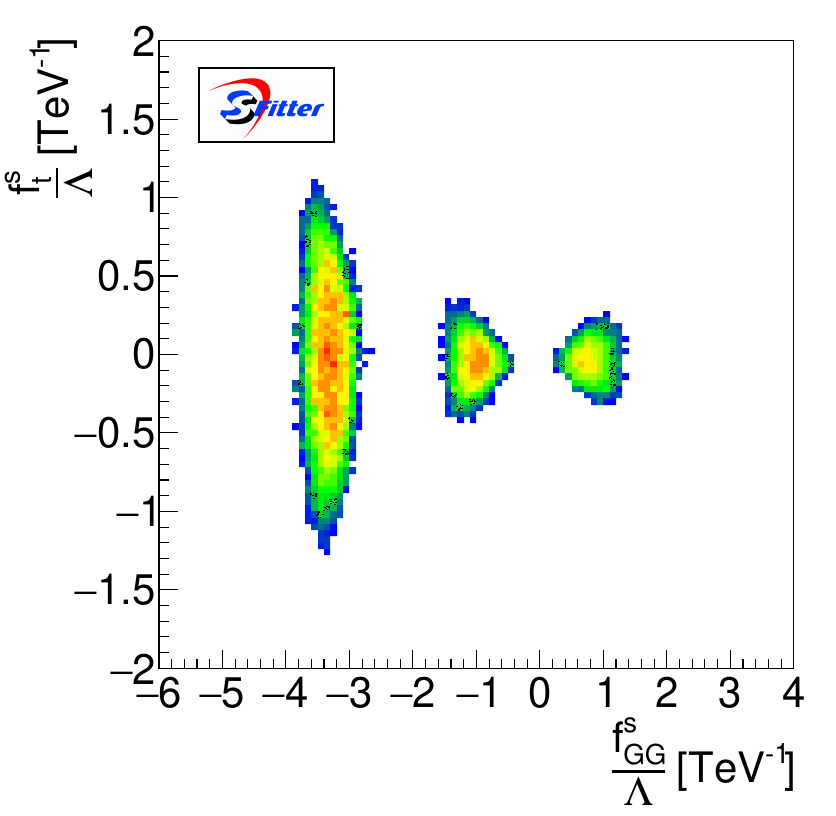}
\hspace{-0.3cm}
\includegraphics[width=0.32\textwidth]{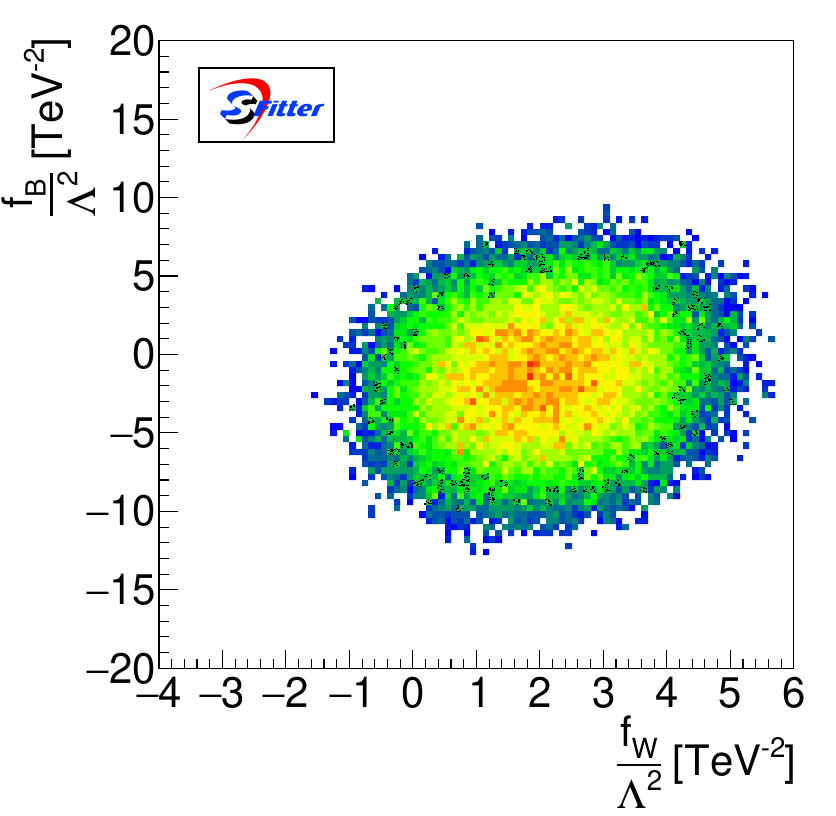}
\hspace{-0.3cm}
\includegraphics[width=0.32\textwidth]{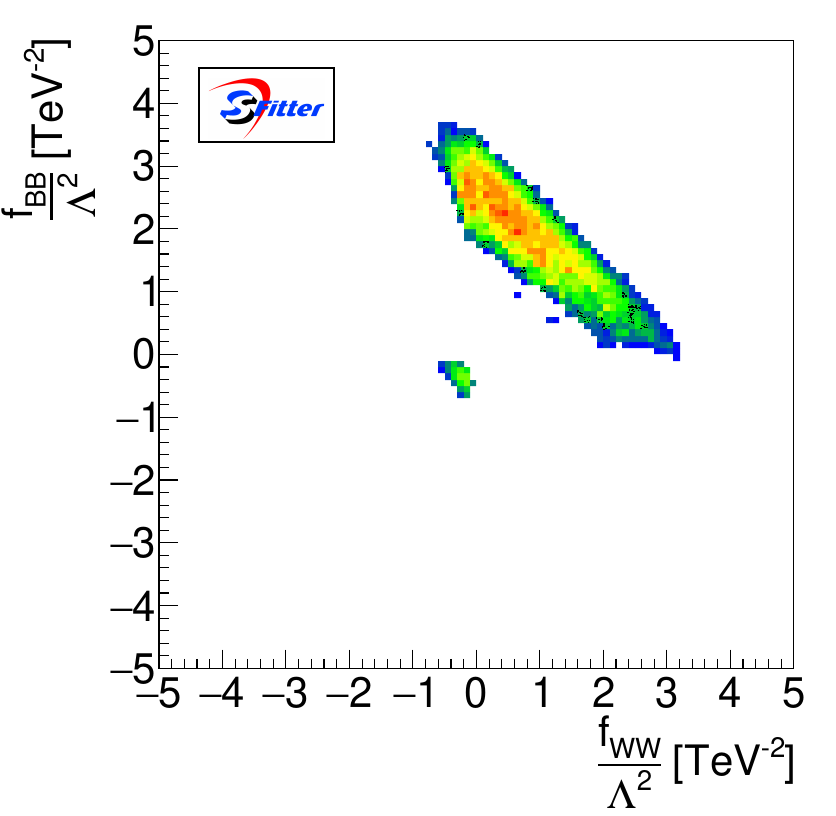}
\hspace{-0.2cm}\raisebox{20pt}{\includegraphics[width=0.047\textwidth]{Figures/barcode.pdf}}
\caption{Two-dimensional profile log-likelihoods for the combined
  Higgs, TGV, and heavy scalar fit, but assuming a common origin of
  operators as defined in Eq.\eqref{eq:prior1}.  The black points
  indicate $\Delta(-2\log L)=5.99$.}
\label{fig:11param_750}
\end{figure}

In Fig.~\ref{fig:11param_750} we again show a selection of
two-dimensional correlations. In the upper-left panel we start with
tight constraints on $f_{BB}^S$ and also on $f_{GG}^S$.  Now
$f_{BB}^S/\Lambda$ no longer presents an unconstrained direction, as
the reduced allowed region for values around $-10~\itev$ is limited from
the constraint that $f_{BB}$ and hence $f_{BB}^S$ is constrained by
the Higgs measurements. Because of the minus signs in
Eq.\eqref{eq:prior1} the region of allowed values for both $f_{BB}^S$
and $f_{WW}^S$ corresponds to the solution that flips the sign of the
$H\gamma\gamma$ vertex while respecting its measured
size~\cite{legacy2}, as seen in the upper-center panel.  In the case
of $f_{GG}^S$ and $f_{GG}$ the several best fit regions are due to the
measurement of a SM--like Higgs boson in gluon fusion production,
the interference between $f_{GG}$ and $f_t$~\cite{legacy1}, and the
heavy scalar anomaly that excludes $f_{GG}$ null values.

As seen in the upper-right panel, the stronger constraints on
$f_{BB}^S$ are directly translated into a stringent 95$\%$ CL bound
on the mixing angle, $\sin\alpha<0.02$. In the lower-left panel we
show the impact of Eq.\eqref{eq:prior1} on $f_t^S$.  The fact that in
this analysis $f_{BB}^S$ is more constrained than in the general
scenario implies that $f_t^S$ is constrained to order one values, as expected from
the $f_{BB}^S$ vs $f_t^S$ correlation in Fig.~\ref{fig:7param_750}.
The solution for $f_t$ that flips the sign of the Higgs-Yukawa present
in Ref.~\cite{legacy1} is excluded through $f_t^S$.  This reduces the
number of allowed regions for $f_{GG}$, as compared to the
electroweak-Higgs fit~\cite{legacy1}.  In the case of $f_\tau^S$ and
$f_b^S$, the allowed regions are limited by the $H\tau^+\tau^-$ and
$Hb\bar{b}$ measurements. The $v/m_f$ factors in Eq.\eqref{eq:prior1}
lead to reduced allowed ranges for $f_\tau^S$ in comparison to the
previous general scenario.

We illustrate in the lower-center panel the allowed region for two of
the dimension--six operators not involved in the simplifications of
Eq.\eqref{eq:prior1}, $f_W$ and $f_B$. They remain unaltered
with respect to the general analysis or the 
electroweak-Higgs results~\cite{legacy2}. Conversely, in the lower-right panel of
Fig.~\ref{fig:11param_750} we illustrate the two parameter regions
$f_{BB}$ vs. $f_{WW}$.
There we see how the SM solution observed in the electroweak-Higgs analysis
is now disfavored with respect to positive values for the Wilson
coefficients.\medskip

In this section we have illustrated the results of a constrained
scenario imposing hard relations between the heavy scalar and Higgs operators 
in Eq.\eqref{eq:prior1}. Realistically, we 
would expect such relations to not be as strict.
We therefore checked that relaxing
Eq.\eqref{eq:prior1} and allowing for order-one variations 
does not qualitatively change our conclusions. Numerically, the bound on the mixing angle $\sin \alpha$ 
becomes weaker once the relation $f_{BB}\propto f_{BB}^S |f_{BB}^S|$ is relaxed. 

\section{Epitaph}
\label{sec:summary}

We have developed the framework to perform a combined analysis of the
electroweak-Higgs sector extended with a new scalar to test Higgs
portal scenarios. The theoretical description we have studied is that
of a linear effective Lagrangian extended with the addition of a
singlet scalar.

The key question we face is the test of the portal structure
hypothesis for an extended scalar sector. With that purpose we include
a large set of Higgs event rates and kinematic distributions, combined
with the recently implemented LHC triple gauge boson vertex
distributions~\cite{legacy2}. As a test of a Higgs portal scenario we
study the possibility that a di-photon anomaly recently observed at
the LHC~\cite{750ex,S750_1000,S750_2100} could be part of an extended
Higgs sector. For that we include a selection of relevant experimental
searches for heavy resonances as listed in
Tab.~\ref{tab:750_data}.\bigskip

Analyzing first the new scalar sector only, we recover the result that
a non-zero value for a reduced set of singlet scalar effective
operators ($f_{GG}^S$, $f_{BB}^S+f_{WW}^S$) fits the observed anomaly
in the di-photon channel, without conflicting with the lack of other
positive observations, see Fig.~\ref{fig:5param_750}. The mixing angle
of the new singlet state with the Higgs boson can be sizable, the
upper bound we find in the analysis is $\sin(\alpha)<0.15$ at the 95\%
CL.  The addition of fermionic dimension--five operators increases the
allowed parameter space regions for the bosonic operators. However it
has no impact on the maximum allowed mixing angle value, see
Fig.~\ref{fig:7param_750}.

We then extend the analysis combining the new scalar sector with the
electroweak-Higgs sector, using the Lagrangian description based on the
dimension--six operators in Eq.\eqref{eq:ourleff}.  In this extended
scenario the upper bound on the mixing angle is further reduced in
order to suppress the new dimension-six contributions to the heavy
scalar non-observed decays.  The upper bound is now $\sin(\alpha)<0.1$
at 95\% CL, with a size still compatible with Higgs portal
hypothesis. Beyond this change, the maximum allowed mixing angle
reduces the correlations between the Higgs-electroweak phenomenology
and the hypothetical heavy scalar interactions. This leads to results
that in the most general scenario are very similar to the ones of the
individual Higgs-electroweak~\cite{legacy2}, and heavy scalar
analysis, respectively.

Motivated by a scalar portal scenario we define and test a hypothesis
for a unique origin of the dimension--five and dimension--six
operators studied in the analysis. Imposing Eq.\eqref{eq:prior1} we
find new physics regions of parameters that fit the di-photon anomaly
while being consistent with the lack of deviations measured on the
electroweak-Higgs measurements. The upper bound on the mixing angle is
reduced in this case to $\sin(\alpha)<0.02$, due to the strong
constraints on the operators modifying $h\rightarrow \gamma\gamma$.

\subsubsection*{Acknowledgments}

MB, AB, JG-F and TP would like to thank the MITP for the hospitality while
this paper was finalized.  They also acknowledge support from the
German Research Foundation (DFG) through the Forschergruppe `New
Physics at the Large Hadron Collider' (FOR 2239), AB also through the
Heidelberg Graduate School for Fundamental Physics and the
Graduiertenkolleg `Particle physics beyond the Standard Model' (GRK
1940).

\appendix
\section*{Higgs-singlet Lagrangian}
\label{sec:app}

We describe here the main details of our effective
Lagrangian analysis. We focus on the Higgs-scalar mixing and the
combined phenomenology we derive. Following the
Lagrangian in Eq.\eqref{eq:lagd5} of Sec.~\ref{sec:lag}, both $\mu_S$
and $f_1^S/\Lambda$ generate a mixing between the two interaction
eigenstates $H^\prime$ and $S^\prime$. In this appendix we denote
interaction eigenstates as primed fields, while mass eigenstates after
the rotation
\begin{align}
\lag_m = 
 -\frac{1}{2} \left(H^\prime\; S^\prime\right) 
\begin{pmatrix}
  M_H^2 & v \left(\mu_S+ \dfrac{f_1^S v^2}{\Lambda} \right)\left(1-\dfrac{f_{\phi,2}v^2}{2 \Lambda^2} \right) \\
 v \left(\mu_S+ \dfrac{f_1^S v^2}{\Lambda} \right)\left(1-\dfrac{f_{\phi,2} v^2}{2 \Lambda^2} \right) & M_S^2 + \dfrac{\lambda_{SH}v}{2} 
\end{pmatrix}
\begin{pmatrix} H^\prime \\ S^\prime \end{pmatrix} \; ,
\label{eq:mixmat}
\end{align}
as un-primed fields. The light mass term is 
$M_H^2 = 2\lambda_H v^2 ( 1- v^2 f_{\phi,2}/\Lambda^2 )$, with the Higgs
quartic $\lambda$. The contribution proportional from $f_{\phi,2}/\Lambda^2$
is originated from the Higgs kinetic term and the appropriate field
re-definition~\cite{barca}. The physical masses are 
\begin{align}
 M_{1,2}^2
= \frac{M_S^2 + \dfrac{\lambda_{SH}v}{2} +M_H^2}{2} \mp \frac{1}{2} \, 
\sqrt{ \left( M_S^2 + \dfrac{\lambda_{SH}v}{2} -M_H^2 \right)^2
      +4v \left(\mu_S+ \frac{f_1^S v^2}{\Lambda} \right)^2
          \left(1-\frac{f_{\phi,2}v^2}{2\Lambda^2} \right)^2} \; ,
\end{align}
and the mixing angle as a function of the physical masses reads
\begin{align}
\sin{2\alpha} 
= \frac{2 v \left(\mu_S+ \dfrac{f_1^S v^2}{\Lambda} \right)
            \left(1-\dfrac{f_{\phi,2}v^2}{2\Lambda^2} \right)}{M_2^2-M_1^2} 
\qquad \stackrel{f_1^S=0}{\Longrightarrow} \qquad 
\mu_S 
= \sin {2\alpha} \, \frac{M_2^2-M_1^2}{2v} \left(1+\frac{v^2}{2}\frac{f_{\phi,2}}{\Lambda^2} \right) \;.
\end{align}
\medskip

The Higgs-scalar mixing affects many couplings of the mass eigenstates $S$ and
$H$.  We first study the interactions of the light, Higgs-like, state.
The admixture of the new scalar generates new interactions of the kind
$s_\alpha f_j^S/\Lambda$, formally of dimension five, with an
additional suppression by the mixing angle. Once we include the
dimension--six operators of $\lag_{\text{dim-6}}^H$, all mixing
contributions can be absorbed in a re-definition of the effective
Higgs Lagrangian, as long as we limit our analysis to tri-linear
interactions. For example, the physical Higgs-gluon coupling becomes
\begin{align}
g_{Hgg} 
= -\frac{\alpha_s}{8 \pi} \left( c_\alpha\frac{f_{GG} v}{\Lambda^2} + 2s_\alpha\frac{f_{GG}^S}{\Lambda}\right)
         \equiv-\frac{\alpha_s}{8 \pi} \frac{f^\prime_{GG} v}{\Lambda^2}
\label{eq:hgg} 
\end{align}
where $g_{Hgg}$ is defined through the term $g_{Hgg} \; H G^a_{\mu\nu}
G^{a\mu\nu}$ in the Lagrangian~\cite{legacy1}. Using these kind of
re-definitions the Higgs part of our analysis can be easily related to
the results of Ref.~\cite{legacy1,legacy2}.

Because the Higgs-scalar mixing of Eq.\eqref{eq:mixmat} is defined in
the broken phase and does not affect the Goldstone modes, this kind of
re-definition does not apply to the triple gauge vertices constrained
by di-boson production channels~\cite{legacy1,legacy2}. The
contribution of $f_W$ and $f_B$ in the Higgs sector is weighted by
$c_\alpha$. For instance, the $f_W$ contribution to the $HWW$
interaction reads
\begin{align}
\lag^{HVV} \supset  g^{(1)}_{H W W} \; \left (W^+_{\mu \nu} W^{- \,
  \mu} \partial^{\nu} H +\text{h.c.} \right)
\qquad \text{with} \qquad 
g^{(1)}_{H W W} 
= c_\alpha \, \frac{g^2 v}{2\Lambda^2} \; \frac{f_W}{2} \; .
\label{eq:hww} 
\end{align}
In contrast, the contributions of $f_W$ and $f_B$ to the triple gauge
boson vertices is not modified by such a mixing angle and remains the
same as in the Higgs--gauge analysis~\cite{legacy2}. This way, a
sizable mixing with the heavy scalar changes the pattern of Higgs-TGV
correlations.\medskip

On the heavy scalar side, the interaction with the incoming gluons is
\begin{align}
g_{Sgg} &=
         -\frac{\alpha_s}{8 \pi} \left(s_\alpha\frac{f_{GG} v}{\Lambda^2} -2c_\alpha\frac{f_{GG}^S}{\Lambda}\right)
         \equiv \frac{\alpha_s}{4\pi}\frac{f_{GG}^{s\prime}}{\Lambda} \;.
\label{eq:sgg}
\end{align}
While the contributions of $f_{WW}\leftrightarrow f_{WW}^S$,
$f_{BB}\leftrightarrow f_{BB}^S$ and the fermionic interactions
$f_{f}\leftrightarrow f_{f}^S$ follow this structure, the case of
$f_W$ and $f_B$ is again special. Both Higgs-like operators generate
new Lorentz structures in the heavy scalar sector.  For example, the
$f_W$ contribution to the $SWW$ vertex is
\begin{align}
\lag^{SVV} \supset g^{(1)}_{S W W}  \; \left (W^+_{\mu \nu} W^{- \, \mu} \partial^{\nu} S 
                            +\text{h.c.} \right)
\qquad \text{with} \qquad 
g^{(1)}_{S W W} 
= s_\alpha \, \frac{g^2 v}{2\Lambda^2} \; \frac{f_W}{2} \; .
\label{eq:hww2} 
\end{align}

Finally, the $SHH$ interaction is generated through the terms
\begin{align}
 \lag \supset \lambda^{SHH}SHH
 +\frac{f_{\phi,2}}{\Lambda^2}s_\alpha c_\alpha^2
 \left(S\partial^\mu H \partial_\mu H + 2 H \partial^\mu S \partial_\mu H \right) \; ,
\end{align}
where the momentum-independent coupling is composed of several terms in
Eq.\eqref{eq:ourleff} and Eq.\eqref{eq:lagd5}
\begin{align}
\lambda^{SHH} =& 
- 3 c_\alpha^2 s_\alpha v \lambda_H \left( 1- \frac{3 f_{\phi,2}v^2}{2\Lambda^2} \right) \notag \\
&+ \frac{1}{2} \left( 2 c_\alpha s_\alpha^2  - c_\alpha^3 \right) 
\mu_S \left( 1-\frac{f_{\phi,2} v^2}{\Lambda^2}\right)
+ \frac{1}{2} \left( 2 c_\alpha^2 s_\alpha - s_\alpha^3 \right) v \lambda_{SH}
  \left( 1-\frac{f_{\phi,2} v^2}{2\Lambda^2}\right) 
- 3 c_\alpha s_\alpha^2 a_3 \notag \\
& + \frac{3}{4} \left( c_\alpha s_\alpha^2 - c_\alpha^3 \right) \frac{f_1^S v^2}{\Lambda} 
    \, \left( 1-\frac{f_{\phi,2} v^2}{\Lambda^2}\right) \notag \\
\equiv&
-3 c_\alpha^2 s_\alpha v \lambda_H \left( 1-\frac{3 f_{\phi,2}v^2}{2\Lambda^2} \right) 
+ \frac{1}{2} \left( 2c_\alpha s_\alpha^2- c_\alpha^3 \right)
  s_{2\alpha} \frac{M_2^2-M_1^2}{v} \left( 1-\frac{f_{\phi,2} v^2}{2\Lambda^2} \right) +c_{SHH}\; .
\label{eq:inthhs}
\end{align}
There the Higgs self coupling $\lambda_H$ can be expressed as
\begin{align}
\lambda_H 
=\frac{s_\alpha^2 M_2^2+c_\alpha^2 M_1^2}{2v^2} \, 
 \left( 1+\frac{f_{\phi,2} v^2}{\Lambda^2}\right) \; ,
\end{align}
while the term $c_{SHH}$ accounts for the contributions from $\lambda_{SH}$,
$a_3$, $f_1^S$ and the dimension-6 operator
$\left(\phi^\dagger\phi\right)^3$.  In a simplified ansatz, we set
$c_{SHH}=0$.



\begin{thebibliography}{99}

\bibitem{higgs}
  P.~W.~Higgs,
  Phys.\ Lett.\  {\bf 12}, 132 (1964);
  P.~W.~Higgs,
  Phys.\ Rev.\ Lett.\  {\bf 13}, 508 (1964);
  F.~Englert and R.~Brout,
  Phys.\ Rev.\ Lett.\  {\bf 13}, 321 (1964).
  
\bibitem{discovery}
  G.~Aad {\it et al.} [ATLAS Collaboration],
  Phys.\ Lett.\ B {\bf 716}, 1 (2012);
  S.~Chatrchyan {\it et al.}  [CMS Collaboration],
  Phys.\ Lett.\ B {\bf 716}, 30 (2012).

\bibitem{barca}
  T.~Corbett, O.~J.~P.~Eboli, J.~Gonzalez-Fraile and M.~C.~Gonzalez-Garcia,
  Phys.\ Rev.\ D {\bf 86}, 075013 (2012);
  T.~Corbett, O.~J.~P.~Eboli, J.~Gonzalez-Fraile and M.~C.~Gonzalez-Garcia,
  Phys.\ Rev.\ D {\bf 87}, 015022 (2013);
  T.~Corbett, O.~J.~P.~Eboli, J.~Gonzalez-Fraile and M.~C.~Gonzalez-Garcia,
  Phys.\ Rev.\ Lett.\  {\bf 111}, 011801 (2013).
  
\bibitem{legacy1}
T.~Corbett, O.~J.~P.~Eboli, D.~Goncalves, J.~Gonzalez-Fraile, T.~Plehn and M.~Rauch,
  JHEP {\bf 1508}, 156 (2015)
  [arXiv:1505.05516 [hep-ph]],
  and [arXiv:1511.08188 [hep-ph]].

\bibitem{legacy2}
  A.~Butter, O.~J.~P.~Eboli, J.~Gonzalez-Fraile, M.~C.~Gonzalez-Garcia, T.~Plehn and M.~Rauch,
  arXiv:1604.03105 [hep-ph].

\bibitem{other}
  E.~Mass\'o and V.~Sanz,
  Phys.\ Rev.\ D {\bf 87}, no. 3, 033001 (2013);
  I.~Brivio, T.~Corbett, O.~J.~P.~Eboli, M.~B.~Gavela, J.~Gonzalez-Fraile, M.~C.~Gonzalez-Garcia, L.~Merlo and S.~Rigolin,
  JHEP {\bf 1403}, 024 (2014);
  G.~Brooijmans {\it et al.},
  arXiv:1405.1617 [hep-ph];
  A.~Falkowski and F.~Riva,
  JHEP {\bf 1502}, 039 (2015).
  A.~Falkowski, M.~Gonzalez-Alonso, A.~Greljo and D.~Marzocca,
  Phys.\ Rev.\ Lett.\  {\bf 116}, no. 1, 011801 (2016);
  I.~Brivio, J.~Gonzalez-Fraile, M.~C.~Gonzalez-Garcia and L.~Merlo,
  arXiv:1604.06801 [hep-ph].
  
\bibitem{portal}
  see \eg 
  B.~Patt and F.~Wilczek,
  hep-ph/0605188;
  R.~Schabinger and J.~D.~Wells,
  Phys.\ Rev.\ D {\bf 72}, 093007 (2005);
  S.~Bock, R.~Lafaye, T.~Plehn, M.~Rauch, D.~Zerwas and P.~M.~Zerwas,
  Phys.\ Lett.\ B {\bf 694}, 44 (2010);
  C.~Englert, T.~Plehn, D.~Zerwas and P.~M.~Zerwas,
  Phys.\ Lett.\ B {\bf 703}, 298 (2011);
  C.~Englert, T.~Plehn, M.~Rauch, D.~Zerwas and P.~M.~Zerwas,
  Phys.\ Lett.\ B {\bf 707}, 512 (2012);
  A.~Djouadi, O.~Lebedev, Y.~Mambrini and J.~Quevillon,
  Phys.\ Lett.\ B {\bf 709}, 65 (2012);
  O.~Lebedev,
  Eur.\ Phys.\ J.\ C {\bf 72}, 2058 (2012).
  
  

\bibitem{750ex}
The ATLAS collaboration,
ATLAS-CONF-2015-081; 
CMS Collaboration,
CMS PAS EXO-15-004.

\bibitem{S750_1000} 
  M.~Aaboud {\it et al.} [ATLAS Collaboration],
  ATLAS-CONF-2016-018,
  arXiv:1606.03833 [hep-ex].

\bibitem{S750_2100} 
  V.~Khachatryan {\it et al.} [CMS Collaboration],
  CMS-PAS-EXO-16-018,
  arXiv:1606.04093 [hep-ex].


\bibitem{not_eft}
  R.~Franceschini {\it et al.},
  JHEP {\bf 1603}, 144 (2016);
  J.~Ellis, S.~A.~R.~Ellis, J.~Quevillon, V.~Sanz and T.~You,
  JHEP {\bf 1603}, 176 (2016);
  D.~Buttazzo, A.~Greljo and D.~Marzocca,
  Eur.\ Phys.\ J.\ C {\bf 76}, no. 3, 116 (2016);
 S.~Knapen, T.~Melia, M.~Papucci and K.~Zurek,
  Phys.\ Rev.\ D {\bf 93}, no. 7, 075020 (2016); 
  S.~D.~McDermott, P.~Meade and H.~Ramani,
  Phys.\ Lett.\ B {\bf 755}, 353 (2016);
  M.~Low, A.~Tesi and L.~T.~Wang,
  JHEP {\bf 1603}, 108 (2016);
  Q.~H.~Cao, Y.~Liu, K.~P.~Xie, B.~Yan and D.~M.~Zhang,
  arXiv:1512.05542 [hep-ph];
  R.~Martinez, F.~Ochoa and C.~F.~Sierra,
  arXiv:1512.05617 [hep-ph];
  D.~Curtin and C.~B.~Verhaaren,
  Phys.\ Rev.\ D {\bf 93}, no. 5, 055011 (2016);
  L.~Bian, N.~Chen, D.~Liu and J.~Shu,
  Phys.\ Rev.\ D {\bf 93}, no. 9, 095011 (2016);
  J.~Chakrabortty, A.~Choudhury, P.~Ghosh, S.~Mondal and T.~Srivastava,
  arXiv:1512.05767 [hep-ph];
  P.~Agrawal, J.~Fan, B.~Heidenreich, M.~Reece and M.~Strassler,
  arXiv:1512.05775 [hep-ph];
  A.~Falkowski, O.~Slone and T.~Volansky,
  JHEP {\bf 1602}, 152 (2016);
  E.~Gabrielli, K.~Kannike, B.~Mele, M.~Raidal, C.~Spethmann and H.~Veerm\"ae
  Phys.\ Lett.\ B {\bf 756}, 36 (2016);
  A.~Alves, A.~G.~Dias and K.~Sinha,
  Phys.\ Lett.\ B {\bf 757}, 39 (2016);
  W.~Chao,
  Phys.\ Rev.\ D {\bf 93}, no. 11, 115013 (2016);
  S.~Chang,
  Phys.\ Rev.\ D {\bf 93}, no. 5, 055016 (2016);
  I.~Chakraborty and A.~Kundu,
  Phys.\ Rev.\ D {\bf 93}, no. 5, 055003 (2016);
  H.~Han, S.~Wang and S.~Zheng,
  Nucl.\ Phys.\ B {\bf 907}, 180 (2016);
  M.~x.~Luo, K.~Wang, T.~Xu, L.~Zhang and G.~Zhu,
  Phys.\ Rev.\ D {\bf 93}, no. 5, 055042 (2016);
  M.~Dhuria and G.~Goswami,
  arXiv:1512.06782 [hep-ph];
  J.~S.~Kim, K.~Rolbiecki and R.~Ruiz de Austri,
  Eur.\ Phys.\ J.\ C {\bf 76}, no. 5, 251 (2016);
  L.~Berthier, J.~M.~Cline, W.~Shepherd and M.~Trott,
  JHEP {\bf 1604}, 084 (2016);
  S.~M.~Boucenna, S.~Morisi and A.~Vicente,
  Phys.\ Rev.\ D {\bf 93}, no. 11, 115008 (2016);
  C.~W.~Murphy,
  Phys.\ Lett.\ B {\bf 757}, 192 (2016);
  A.~E.~C.~Hernandez and I.~Nisandzic,
  arXiv:1512.07165 [hep-ph];
  J.~de Blas, J.~Santiago and R.~Vega-Morales,
  Phys.\ Lett.\ B {\bf 759}, 247 (2016);
  K.~M.~Patel and P.~Sharma,
  Phys.\ Lett.\ B {\bf 757}, 282 (2016);
  J.~Gu and Z.~Liu,
  Phys.\ Rev.\ D {\bf 93}, no. 7, 075006 (2016);
  K.~Das and S.~K.~Rai,
  Phys.\ Rev.\ D {\bf 93}, no. 9, 095007 (2016);
  J.~Zhang and S.~Zhou,
  Chin.\ Phys.\ C {\bf 40}, 081001 (2016);
  D.~Chway, R.~Dermisek, T.~H.~Jung and H.~D.~Kim,
  arXiv:1512.08221 [hep-ph];
  G.~Li, Y.~n.~Mao, Y.~L.~Tang, C.~Zhang, Y.~Zhou and S.~h.~Zhu,
  Phys.\ Rev.\ Lett.\  {\bf 116}, no. 15, 151803 (2016);
  H.~An, C.~Cheung and Y.~Zhang,
  arXiv:1512.08378 [hep-ph];
  Q.~H.~Cao, Y.~Liu, K.~P.~Xie, B.~Yan and D.~M.~Zhang,
  Phys.\ Rev.\ D {\bf 93}, no. 7, 075030 (2016);
  J.~Cao, L.~Shang, W.~Su, F.~Wang and Y.~Zhang,
  arXiv:1512.08392 [hep-ph];
  J.~Gao, H.~Zhang and H.~X.~Zhu,
  arXiv:1512.08478 [hep-ph];
  F.~Goertz, J.~F.~Kamenik, A.~Katz and M.~Nardecchia,
  JHEP {\bf 1605}, 187 (2016);
  C.~W.~Chiang, M.~Ibe and T.~T.~Yanagida,
  JHEP {\bf 1605}, 084 (2016);
  S.~Kanemura, N.~Machida, S.~Odori and T.~Shindou,
  arXiv:1512.09053 [hep-ph];
  I.~Low and J.~Lykken,
  arXiv:1512.09089 [hep-ph];
  K.~Kaneta, S.~Kang and H.~S.~Lee,
  arXiv:1512.09129 [hep-ph];
  A.~Dasgupta, M.~Mitra and D.~Borah,
  arXiv:1512.09202 [hep-ph].
  

  
  
  
  
  

\bibitem{not_vectors} 
  A.~Pilaftsis,
  Phys.\ Rev.\ D {\bf 93}, no. 1, 015017 (2016);
  S.~Di Chiara, L.~Marzola and M.~Raidal,
  Phys.\ Rev.\ D {\bf 93}, no. 9, 095018 (2016);
  R.~S.~Gupta, S.~Jager, Y.~Kats, G.~Perez and E.~Stamou,
  arXiv:1512.05332 [hep-ph];
  A.~Kobakhidze, F.~Wang, L.~Wu, J.~M.~Yang and M.~Zhang,
  Phys.\ Lett.\ B {\bf 757}, 92 (2016);
  Y.~Bai, J.~Berger and R.~Lu,
  Phys.\ Rev.\ D {\bf 93}, no. 7, 076009 (2016);
  J.~Chang, K.~Cheung and C.~T.~Lu,
  Phys.\ Rev.\ D {\bf 93}, no. 7, 075013 (2016);
  O.~Antipin, M.~Mojaza and F.~Sannino,
  Phys.\ Rev.\ D {\bf 93}, no. 11, 115007 (2016);
  J.~Cao, C.~Han, L.~Shang, W.~Su, J.~M.~Yang and Y.~Zhang,
  Phys.\ Lett.\ B {\bf 755}, 456 (2016);
  U.~K.~Dey, S.~Mohanty and G.~Tomar,
  Phys.\ Lett.\ B {\bf 756}, 384 (2016);
  G.~M.~Pelaggi, A.~Strumia and E.~Vigiani,
  JHEP {\bf 1603}, 025 (2016);
  W.~Altmannshofer, J.~Galloway, S.~Gori, A.~L.~Kagan, A.~Martin and J.~Zupan,
  Phys.\ Rev.\ D {\bf 93}, no. 9, 095015 (2016);
  H.~Davoudiasl and C.~Zhang,
  Phys.\ Rev.\ D {\bf 93}, no. 5, 055006 (2016);
  N.~Craig, P.~Draper, C.~Kilic and S.~Thomas,
  Phys.\ Rev.\ D {\bf 93}, 115023 (2016);
  J.~E.~Kim,
  Phys.\ Lett.\ B {\bf 755}, 190 (2016);
  W.~Chao,
  arXiv:1512.08484 [hep-ph];
  X.~J.~Bi {\it et al.},
  arXiv:1512.08497 [hep-ph];
  P.~S.~B.~Dev, R.~N.~Mohapatra and Y.~Zhang,
  JHEP {\bf 1602}, 186 (2016);
  S.~Kanemura, K.~Nishiwaki, H.~Okada, Y.~Orikasa, S.~C.~Park and R.~Watanabe,
  arXiv:1512.09048 [hep-ph];
  A.~E.~C.~Hernandez,
  arXiv:1512.09092 [hep-ph];
  L.~Marzola, A.~Racioppi, M.~Raidal, F.~R.~Urban and H.~Veerm\"ae,
  JHEP {\bf 1603}, 190 (2016);
  

\bibitem{not_susy}
  W.~Chao, R.~Huo and J.~H.~Yu,
  arXiv:1512.05738 [hep-ph];
  S.~Chakraborty, A.~Chakraborty and S.~Raychaudhuri,
  arXiv:1512.07527 [hep-ph];
  Y.~L.~Tang and S.~h.~Zhu,
  arXiv:1512.08323 [hep-ph];
  B.~Dutta, Y.~Gao, T.~Ghosh, I.~Gogoladze, T.~Li, Q.~Shafi and J.~W.~Walker,
  arXiv:1601.00866 [hep-ph];
  L.~M.~Carpenter, R.~Colburn and J.~Goodman,
  arXiv:1512.06107 [hep-ph];
  H.~P.~Nilles and M.~W.~Winkler,
  JHEP {\bf 1605}, 182 (2016);
  M.~Badziak, M.~Olechowski, S.~Pokorski and K.~Sakurai,
  arXiv:1603.02203 [hep-ph];
  G.~Lazarides and Q.~Shafi,
  Phys.\ Rev.\ D {\bf 93}, 111702 (2016);
  C.~Han, T.~T.~Yanagida and N.~Yokozaki,
  Phys.\ Rev.\ D {\bf 93}, no. 5, 055025 (2016);
  S.~F.~King and R.~Nevzorov,
  JHEP {\bf 1603}, 139 (2016);
  W.~Chao,
  arXiv:1601.00633 [hep-ph];
  L.~J.~Hall, K.~Harigaya and Y.~Nomura,
  JHEP {\bf 1603}, 017 (2016);
  B.~C.~Allanach, P.~S.~B.~Dev, S.~A.~Renner and K.~Sakurai,
  Phys.\ Rev.\ D {\bf 93}, 115022 (2016);
  R.~Ding, L.~Huang, T.~Li and B.~Zhu,
  arXiv:1512.06560 [hep-ph];
  T.~F.~Feng, X.~Q.~Li, H.~B.~Zhang and S.~M.~Zhao,
  arXiv:1512.06696 [hep-ph];
  U.~Ellwanger and C.~Hugonie,
  JHEP {\bf 1605}, 114 (2016);
  T.~Cohen, G.~D.~Kribs, A.~E.~Nelson and B.~Ostdiek,
  arXiv:1605.04308 [hep-ph].
  D.~M.~Ghilencea and H.~M.~Lee,
  arXiv:1606.04131 [hep-ph].
  H.~K.~Dreiner, M.~E.~Krauss, B.~O'Leary, T.~Opferkuch and F.~Staub,
  arXiv:1606.08811 [hep-ph].
  
\bibitem{not_sgoldstino}
  C.~Petersson and R.~Torre,
  Phys.\ Rev.\ Lett.\  {\bf 116}, no. 15, 151804 (2016);
  S.~V.~Demidov and D.~S.~Gorbunov,
  JETP Lett.\  {\bf 103}, no. 4, 219 (2016);
  J.~A.~Casas, J.~R.~Espinosa and J.~M.~Moreno,
  Phys.\ Lett.\ B {\bf 759}, 159 (2016);
  P.~Baratella, J.~Elias-Miro, J.~Penedo and A.~Romanino,
  arXiv:1603.05682 [hep-ph];
  R.~Ding, Y.~Fan, L.~Huang, C.~Li, T.~Li, S.~Raza and B.~Zhu,
  arXiv:1602.00977 [hep-ph].

\bibitem{not_dilaton}
  J.~Cao, L.~Shang, W.~Su, Y.~Zhang and J.~Zhu,
  Eur.\ Phys.\ J.\ C {\bf 76}, no. 5, 239 (2016);
  E.~Megias, O.~Pujolas and M.~Quiros,
  JHEP {\bf 1605}, 137 (2016);
  B.~Agarwal, J.~Isaacson and K.~A.~Mohan,
  arXiv:1604.05328 [hep-ph];
  D.~K.~Hong and D.~H.~Kim,
  Phys.\ Lett.\ B {\bf 758}, 370 (2016).
  
\bibitem{not_radion}
  P.~Cox, A.~D.~Medina, T.~S.~Ray and A.~Spray,
  arXiv:1512.05618 [hep-ph];
  A.~Ahmed, B.~M.~Dillon, B.~Grzadkowski, J.~F.~Gunion and Y.~Jiang,
  arXiv:1512.05771 [hep-ph];
  E.~E.~Boos, V.~E.~Bunichev and I.~P.~Volobuev,
  arXiv:1603.04495 [hep-ph];
  D.~Bardhan, D.~Bhatia, A.~Chakraborty, U.~Maitra, S.~Raychaudhuri and T.~Samui,
  arXiv:1512.06674 [hep-ph].
  M.~Frank, K.~Huitu, U.~Maitra and M.~Patra,
  arXiv:1606.07689 [hep-ph].
  F.~Abu-Ajamieh, R.~Houtz and R.~Zheng,
  arXiv:1607.01464 [hep-ph].


  
\bibitem{not_extrad1}
  C.~Cai, Z.~H.~Yu and H.~H.~Zhang,
  Phys.\ Rev.\ D {\bf 93}, no. 7, 075033 (2016);
  S.~Abel and V.~V.~Khoze,
  JHEP {\bf 1605}, 063 (2016).

\bibitem{not_extrad2}
  M.~Bauer, C.~Hoerner and M.~Neubert,
  arXiv:1603.05978 [hep-ph];
  C.~Csaki and L.~Randall,
  arXiv:1603.07303 [hep-ph].
  
\bibitem{not_composite}
  K.~Harigaya and Y.~Nomura,
  JHEP {\bf 1603}, 091 (2016);
  D.~B.~Franzosi and M.~T.~Frandsen,
  arXiv:1601.05357 [hep-ph];
  M.~Son and A.~Urbano,
  JHEP {\bf 1605}, 181 (2016);
  A.~Belyaev, G.~Cacciapaglia, H.~Cai, T.~Flacke, A.~Parolini and H.~Seradio,
  arXiv:1512.07242 [hep-ph];
  Y.~Nakai, R.~Sato and K.~Tobioka,
  Phys.\ Rev.\ Lett.\  {\bf 116}, no. 15, 151802 (2016);
  J.~M.~No, V.~Sanz and J.~Setford,
  Phys.\ Rev.\ D {\bf 93}, no. 9, 095010 (2016);
  E.~Molinaro, F.~Sannino and N.~Vignaroli,
  arXiv:1512.05334 [hep-ph];
  N.~D.~Barrie, A.~Kobakhidze, M.~Talia and L.~Wu,
  Phys.\ Lett.\ B {\bf 755}, 343 (2016);
  E.~Molinaro, F.~Sannino and N.~Vignaroli,
  arXiv:1602.07574 [hep-ph];
  M.~Redi, A.~Strumia, A.~Tesi and E.~Vigiani,
  JHEP {\bf 1605}, 078 (2016);
  K.~Harigaya and Y.~Nomura,
  arXiv:1603.05774 [hep-ph];
  P.~Ko, C.~Yu and T.~C.~Yuan,
  arXiv:1603.08802 [hep-ph];
  B.~Bellazzini, R.~Franceschini, F.~Sala and J.~Serra,
  JHEP {\bf 1604}, 072 (2016);
  P.~Lebiedowicz, M.~Luszczak, R.~Pasechnik and A.~Szczurek,
  arXiv:1604.02037 [hep-ph];
  W.~Liao and H.~q.~Zheng,
  arXiv:1512.06741 [hep-ph];
  J.~M.~Cline and Z.~Liu,
  arXiv:1512.06827 [hep-ph];
  C.~Han, K.~Ichikawa, S.~Matsumoto, M.~M.~Nojiri and M.~Takeuchi,
  JHEP {\bf 1604}, 159 (2016).
  
\bibitem{not_spin2}
  M.~T.~Arun and P.~Saha,
  arXiv:1512.06335 [hep-ph];
  C.~Han, H.~M.~Lee, M.~Park and V.~Sanz,
  Phys.\ Lett.\ B {\bf 755}, 371 (2016);
  U.~Danielsson, R.~Enberg, G.~Ingelman and T.~Mandal,
  arXiv:1601.00624 [hep-ph];
  C.~Q.~Geng and D.~Huang,
  arXiv:1601.07385 [hep-ph];
  S.~B.~Giddings and H.~Zhang,
  Phys.\ Rev.\ D {\bf 93}, no. 11, 115002 (2016);
  A.~Falkowski and J.~F.~Kamenik,
  arXiv:1603.06980 [hep-ph];
  A.~Carmona,
  arXiv:1603.08913 [hep-ph];
  B.~M.~Dillon and V.~Sanz,
  arXiv:1603.09550 [hep-ph];
  J.~L.~Hewett and T.~G.~Rizzo,
  arXiv:1603.08250 [hep-ph].
  E.~Alvarez, L.~Da Rold, J.~Mazzitelli and A.~Szynkman,
  arXiv:1606.05326 [hep-ph].
  A.~Kobakhidze, K.~McDonald, L.~Wu and J.~Yue,
  arXiv:1606.08565 [hep-ph].


\bibitem{not_interference1}
  J.~H.~Davis, M.~Fairbairn, J.~Heal and P.~Tunney,
  arXiv:1601.03153 [hep-ph];
  B.~J.~Kavanagh,
  arXiv:1601.07330 [hep-ph].

\bibitem{not_wide}
  M.~R.~Buckley,
  arXiv:1601.04751 [hep-ph];
  D.~Aloni, K.~Blum, A.~Dery, A.~Efrati and Y.~Nir,
  arXiv:1512.05778 [hep-ph];
  A.~Salvio, F.~Staub, A.~Strumia and A.~Urbano,
  JHEP {\bf 1603}, 214 (2016).

\bibitem{not_interference2}
  S.~Jung, J.~Song and Y.~W.~Yoon,
  JHEP {\bf 1605}, 009 (2016);
  A.~Djouadi, J.~Ellis and J.~Quevillon,
  arXiv:1605.00542 [hep-ph];
  S.~P.~Martin,
  arXiv:1606.03026 [hep-ph].
  B.~Hespel, F.~Maltoni and E.~Vryonidou,
  arXiv:1606.04149 [hep-ph].
  
\bibitem{not_dm}  
  M.~Backovic, A.~Mariotti and D.~Redigolo,
  JHEP {\bf 1603}, 157 (2016);
  F.~D'Eramo, J.~de Vries and P.~Panci,
  JHEP {\bf 1605}, 089 (2016);
  M.~T.~Frandsen and I.~M.~Shoemaker,
  JCAP {\bf 1605}, no. 05, 064 (2016);
  G.~Arcadi, P.~Ghosh, Y.~Mambrini and M.~Pierre,
  arXiv:1603.05601 [hep-ph];
  E.~Morgante, D.~Racco, M.~Rameez and A.~Riotto,
  arXiv:1603.05592 [hep-ph];
  S.~F.~Ge, H.~J.~He, J.~Ren and Z.~Z.~Xianyu,
  Phys.\ Lett.\ B {\bf 757}, 480 (2016);
  A.~Hektor and L.~Marzola,
  arXiv:1602.00004 [hep-ph];
  Q.~H.~Cao, Y.~Q.~Gong, X.~Wang, B.~Yan and L.~L.~Yang,
  Phys.\ Rev.\ D {\bf 93}, no. 7, 075034 (2016);
  P.~Ko and T.~Nomura,
  Phys.\ Lett.\ B {\bf 758}, 205 (2016);
  S.~Bhattacharya, S.~Patra, N.~Sahoo and N.~Sahu,
  JCAP {\bf 1606}, no. 06, 010 (2016);
  A.~Berlin,
  Phys.\ Rev.\ D {\bf 93}, no. 5, 055015 (2016);
  X.~J.~Huang, W.~H.~Zhang and Y.~F.~Zhou,
  Phys.\ Rev.\ D {\bf 93}, 115006 (2016);
  J.~C.~Park and S.~C.~Park,
  arXiv:1512.08117 [hep-ph];
  H.~Han, S.~Wang and S.~Zheng,
  arXiv:1512.07992 [hep-ph];
  P.~S.~B.~Dev and D.~Teresi,
  arXiv:1512.07243 [hep-ph];
  X.~J.~Bi, Q.~F.~Xiang, P.~F.~Yin and Z.~H.~Yu,
  Nucl.\ Phys.\ B {\bf 909}, 43 (2016);
  D.~Barducci, A.~Goudelis, S.~Kulkarni and D.~Sengupta,
  JHEP {\bf 1605}, 154 (2016);
  K.~Ghorbani and H.~Ghorbani,
  arXiv:1601.00602 [hep-ph];
  S.~M.~Choi, Y.~J.~Kang and H.~M.~Lee,
  arXiv:1605.04804 [hep-ph].
  
\bibitem{not_exotics}
  J.~S.~Kim, J.~Reuter, K.~Rolbiecki and R.~Ruiz de Austri,
  Phys.\ Lett.\ B {\bf 755}, 403 (2016);
  J.~Bernon and C.~Smith,
  Phys.\ Lett.\ B {\bf 757}, 148 (2016);
  W.~S.~Cho, D.~Kim, K.~Kong, S.~H.~Lim, K.~T.~Matchev, J.~C.~Park and M.~Park,
  Phys.\ Rev.\ Lett.\  {\bf 116}, no. 15, 151805 (2016);
  F.~P.~Huang, C.~S.~Li, Z.~L.~Liu and Y.~Wang,
  arXiv:1512.06732 [hep-ph];
  M.~Chala, M.~Duerr, F.~Kahlhoefer and K.~Schmidt-Hoberg,
  Phys.\ Lett.\ B {\bf 755}, 145 (2016);
  J.~Liu, X.~P.~Wang and W.~Xue,
  arXiv:1512.07885 [hep-ph];
  X.~F.~Han, L.~Wang, L.~Wu, J.~M.~Yang and M.~Zhang,
  Phys.\ Lett.\ B {\bf 756}, 309 (2016);
  V.~De Romeri, J.~S.~Kim, V.~Martin-Lozano, K.~Rolbiecki and R.~R.~de Austri,
  Eur.\ Phys.\ J.\ C {\bf 76}, no. 5, 262 (2016);
  B.~A.~Dobrescu, P.~J.~Fox and J.~Kearney,
  arXiv:1605.08772 [hep-ph].
  B.~M.~Dillon, C.~Han, H.~M.~Lee and M.~Park,
  arXiv:1606.07171 [hep-ph].

\bibitem{not_photon}
  H.~Ito, T.~Moroi and Y.~Takaesu,
  Phys.\ Lett.\ B {\bf 756}, 147 (2016);
  M.~He, X.~G.~He and Y.~Tang,
  Phys.\ Lett.\ B {\bf 759}, 166 (2016);
  S.~Fichet, G.~von Gersdorff and C.~Royon,
  Phys.\ Rev.\ D {\bf 93}, no. 7, 075031 (2016);
  C.~Csaki, J.~Hubisz, S.~Lombardo and J.~Terning,
  Phys.\ Rev.\ D {\bf 93}, no. 9, 095020 (2016);
  L.~A.~Harland-Lang, V.~A.~Khoze and M.~G.~Ryskin,
  Eur.\ Phys.\ J.\ C {\bf 76}, no. 5, 255 (2016);
  L.~A.~Harland-Lang, V.~A.~Khoze and M.~G.~Ryskin,
  JHEP {\bf 1603}, 182 (2016);
  C.~Csaki, J.~Hubisz and J.~Terning,
  Phys.\ Rev.\ D {\bf 93}, no. 3, 035002 (2016);
  A.~D.~Martin and M.~G.~Ryskin,
  J.\ Phys.\ G {\bf 43}, no. 4, 04LT02 (2016);
  F.~Richard,
  arXiv:1604.01640 [hep-ex].
  
\bibitem{not_flavor}
  D.~Buttazzo, A.~Greljo, G.~Isidori and D.~Marzocca,
  arXiv:1604.03940 [hep-ph];
  F.~F.~Deppisch, S.~Kulkarni, H.~P\"as and E.~Schumacher,
  arXiv:1603.07672 [hep-ph];
  G.~Belanger and C.~Delaunay,
  arXiv:1603.03333 [hep-ph];
  M.~Bauer and M.~Neubert,
  arXiv:1512.06828 [hep-ph].

\bibitem{not_phase}
  M.~Perelstein and Y.~D.~Tsai,
  arXiv:1603.04488 [hep-ph];
  W.~Chao,
  arXiv:1601.04678 [hep-ph];
  A.~Ghoshal,
  arXiv:1601.04291 [hep-ph].
  
\bibitem{not_strong}
  Q.~H.~Cao, S.~L.~Chen and P.~H.~Gu,
  arXiv:1512.07541 [hep-ph];
  P.~Draper and D.~McKeen,
  JHEP {\bf 1604}, 127 (2016);
  C.~W.~Chiang, H.~Fukuda, M.~Ibe and T.~T.~Yanagida,
  Phys.\ Rev.\ D {\bf 93}, no. 9, 095016 (2016);
  T.~Higaki, K.~S.~Jeong, N.~Kitajima and F.~Takahashi,
  Phys.\ Lett.\ B {\bf 755}, 13 (2016);
  S.~Dimopoulos, A.~Hook, J.~Huang and G.~Marques-Tavares,
  arXiv:1606.03097 [hep-ph].

\bibitem{not_strings}  
  B.~Dutta, Y.~Gao, T.~Ghosh, I.~Gogoladze and T.~Li,
  Phys.\ Rev.\ D {\bf 93}, no. 5, 055032 (2016);
  L.~A.~Anchordoqui, I.~Antoniadis, H.~Goldberg, X.~Huang, D.~Lust and T.~R.~Taylor,
  Phys.\ Lett.\ B {\bf 755}, 312 (2016);
  T.~Li, J.~A.~Maxin, V.~E.~Mayes and D.~V.~Nanopoulos,
  arXiv:1602.09099 [hep-ph];
  L.~A.~Anchordoqui, I.~Antoniadis, H.~Goldberg, X.~Huang, D.~Lust and T.~R.~Taylor,
  Phys.\ Lett.\ B {\bf 759}, 223 (2016);
  G.~K.~Leontaris and Q.~Shafi,
  arXiv:1603.06962 [hep-ph];
  M.~Cvetic, J.~Halverson and P.~Langacker,
  arXiv:1602.06257 [hep-ph];
  A.~E.~Faraggi and J.~Rizos,
  Eur.\ Phys.\ J.\ C {\bf 76}, no. 3, 170 (2016);
  A.~Karozas, S.~F.~King, G.~K.~Leontaris and A.~K.~Meadowcroft,
  Phys.\ Lett.\ B {\bf 757}, 73 (2016);
  E.~Palti,
  Nucl.\ Phys.\ B {\bf 907}, 597 (2016);
  M.~Cvetic, J.~Halverson and P.~Langacker,
  arXiv:1512.07622 [hep-ph];
  J.~J.~Heckman,
  Nucl.\ Phys.\ B {\bf 906}, 231 (2016);
  L.~E.~Ibanez and V.~Martin-Lozano,
  arXiv:1512.08777 [hep-ph];
  J.~Ashfaque, L.~Delle Rose, A.~E.~Faraggi and C.~Marzo,
  arXiv:1606.01052 [hep-ph].
  A.~Belhaj and S.~E.~Ennadifi,
  arXiv:1606.02956 [hep-ph].

\bibitem{not_stuff}
  running out of steam categorizing all papers on the 750~GeV resonance 
  we refer to more interesting aspects in
  T.~Nomura and H.~Okada,
  Phys.\ Lett.\ B {\bf 755}, 306 (2016);
  P.~Ko, Y.~Omura and C.~Yu,
  JHEP {\bf 1604}, 098 (2016);
  A.~E.~C.~Hernandez, I.~d.~M.~Varzielas and E.~Schumacher,
  arXiv:1601.00661 [hep-ph];
  R.~Barbieri, D.~Buttazzo, L.~J.~Hall and D.~Marzocca,
  arXiv:1603.00718 [hep-ph];
  T.~Modak, S.~Sadhukhan and R.~Srivastava,
  Phys.\ Lett.\ B {\bf 756}, 405 (2016);
  F.~F.~Deppisch, C.~Hati, S.~Patra, P.~Pritimita and U.~Sarkar,
  Phys.\ Lett.\ B {\bf 757}, 223 (2016);
  H.~Zhang,
  arXiv:1601.01355 [hep-ph];
  I.~Sahin,
  arXiv:1601.01676 [hep-ph];
  D.~Borah, S.~Patra and S.~Sahoo,
  arXiv:1601.01828 [hep-ph];
  D.~Stolarski and R.~Vega-Morales,
  Phys.\ Rev.\ D {\bf 93}, no. 5, 055008 (2016);
  M.~Fabbrichesi and A.~Urbano,
  arXiv:1601.02447 [hep-ph];
  C.~Hati,
  Phys.\ Rev.\ D {\bf 93}, no. 7, 075002 (2016);
  J.~H.~Yu,
  Phys.\ Rev.\ D {\bf 93}, no. 11, 113007 (2016);
  R.~Ding, Z.~L.~Han, Y.~Liao and X.~D.~Ma,
  Eur.\ Phys.\ J.\ C {\bf 76}, no. 4, 204 (2016);
  I.~Dorsner, S.~Fajfer and N.~Kosnik,
  arXiv:1601.03267 [hep-ph];
  T.~Nomura and H.~Okada,
  arXiv:1601.04516 [hep-ph];
  H.~Okada and K.~Yagyu,
  Phys.\ Lett.\ B {\bf 756}, 337 (2016);
  U.~Aydemir and T.~Mandal,
  arXiv:1601.06761 [hep-ph];
  A.~Djouadi, J.~Ellis, R.~Godbole and J.~Quevillon,
  JHEP {\bf 1603}, 205 (2016);
  T.~Nomura and H.~Okada,
  Phys.\ Lett.\ B {\bf 756}, 295 (2016);
  E.~Bertuzzo, P.~A.~N.~Machado and M.~Taoso,
  arXiv:1601.07508 [hep-ph];
  J.~Kawamura and Y.~Omura,
  Phys.\ Rev.\ D {\bf 93}, no. 11, 115011 (2016);
  I.~Ben-Dayan and R.~Brustein,
  arXiv:1601.07564 [hep-ph];
  L.~Aparicio, A.~Azatov, E.~Hardy and A.~Romanino,
  JHEP {\bf 1605}, 077 (2016);
  T.~Li, J.~A.~Maxin, V.~E.~Mayes and D.~V.~Nanopoulos,
  arXiv:1602.01377 [hep-ph];
  S.~I.~Godunov, A.~N.~Rozanov, M.~I.~Vysotsky and E.~V.~Zhemchugov,
  arXiv:1602.02380 [hep-ph];
  C.~Arbelez, A.~E.~C.~Hernandez, S.~Kovalenko and I.~Schmidt,
  arXiv:1602.03607 [hep-ph];
  C.~Gross, O.~Lebedev and J.~M.~No,
  arXiv:1602.03877 [hep-ph];
  Y.~Hamada, H.~Kawai, K.~Kawana and K.~Tsumura,
  arXiv:1602.04170 [hep-ph];
  F.~Goertz, A.~Katz, M.~Son and A.~Urbano,
  arXiv:1602.04801 [hep-ph];
  C.~Delaunay and Y.~Soreq,
  arXiv:1602.04838 [hep-ph];
  S.~F.~Mantilla, R.~Martinez, F.~Ochoa and C.~F.~Sierra,
  arXiv:1602.05216 [hep-ph];
  Y.~J.~Zhang, B.~B.~Zhou and J.~J.~Sun,
  arXiv:1602.05539 [hep-ph];
  F.~Staub {\it et al.},
  arXiv:1602.05581 [hep-ph];
  S.~Baek and J.~h.~Park,
  Phys.\ Lett.\ B {\bf 758}, 416 (2016);
  P.~Ko, T.~Nomura, H.~Okada and Y.~Orikasa,
  arXiv:1602.07214 [hep-ph];
  J.~Ren and J.~H.~Yu,
  arXiv:1602.07708 [hep-ph];
  J.~F.~Kamenik and M.~Redi,
  arXiv:1603.07719 [hep-ph];
  Y.~Kats and M.~J.~Strassler,
  JHEP {\bf 1605}, 092 (2016);
  A.~Ahriche, G.~Faisel, S.~Nasri and J.~Tandean,
  arXiv:1603.01606 [hep-ph];
  J.~Bernon, A.~Goudelis, S.~Kraml, K.~Mawatari and D.~Sengupta,
  JHEP {\bf 1605}, 128 (2016);
  G.~Panico, L.~Vecchi and A.~Wulzer,
  arXiv:1603.04248 [hep-ph];
  A.~Bharucha, A.~Djouadi and A.~Goudelis,
  arXiv:1603.04464 [hep-ph].
  W.~Lu,
  arXiv:1603.04697 [physics.gen-ph];
  D.~T.~Huong and P.~V.~Dong,
  Phys.\ Rev.\ D {\bf 93}, no. 9, 095019 (2016);
  J.~F.~Kamenik, B.~R.~Safdi, Y.~Soreq and J.~Zupan,
  arXiv:1603.06566 [hep-ph];
  X.~Liu and H.~Zhang,
  arXiv:1603.07190 [hep-ph];
  S.~Di Chiara, A.~Hektor, K.~Kannike, L.~Marzola and M.~Raidal,
  arXiv:1603.07263 [hep-ph];
  K.~Howe, S.~Knapen and D.~J.~Robinson,
  arXiv:1603.08932 [hep-ph];
  J.~H.~Collins, C.~Csaki, J.~A.~Dror and S.~Lombardo,
  Phys.\ Rev.\ D {\bf 93}, no. 11, 115001 (2016);
  N.~Liu, W.~Wang, M.~Zhang and R.~Zheng,
  arXiv:1604.00728 [hep-ph];
  G.~Cynolter, J.~.Kovacs and E.~Lendvai,
  arXiv:1604.01008 [hep-ph];
  T.~Gherghetta, N.~Nagata and M.~Shifman,
  Phys.\ Rev.\ D {\bf 93}, no. 11, 115010 (2016);
  M.~Chala, C.~Grojean, M.~Riembau and T.~Vantalon,
  arXiv:1604.02029 [hep-ph];
  R.~S.~Chivukula, A.~Farzinnia, K.~Mohan and E.~H.~Simmons,
  arXiv:1604.02157 [hep-ph];
  A.~Y.~Kamenshchik, A.~A.~Starobinsky, A.~Tronconi, G.~P.~Vacca and G.~Venturi,
  arXiv:1604.02371 [hep-ph];
  N.~D.~Barrie, A.~Kobakhidze, S.~Liang, M.~Talia and L.~Wu,
  arXiv:1604.02803 [hep-ph];
  A.~Kusenko, L.~Pearce and L.~Yang,
  Phys.\ Rev.\ D {\bf 93}, no. 11, 115005 (2016);
  H.~Ito and T.~Moroi,
  arXiv:1604.04076 [hep-ph];
  A.~Bolanos, J.~L.~Diaz-Cruz, G.~Hernandez-Tome and G.~Tavares-Velasco,
  arXiv:1604.04822 [hep-ph];
  R.~Franceschini, G.~F.~Giudice, J.~F.~Kamenik, M.~McCullough, F.~Riva, A.~Strumia and R.~Torre,
  arXiv:1604.06446 [hep-ph];
  M.~Duerr, P.~Fileviez Perez and J.~Smirnov,
  arXiv:1604.05319 [hep-ph];
  S.~Iwamoto, G.~Lee, Y.~Shadmi and R.~Ziegler,
  arXiv:1604.07776 [hep-ph].
  S.~Fichet, G.~von Gersdorff and C.~Royon,
  Phys.\ Rev.\ Lett.\  {\bf 116}, no. 23, 231801 (2016);
  Y.~Cai, J.~D.~Clarke, R.~R.~Volkas and T.~T.~Yanagida,
  arXiv:1605.02743 [hep-ph];
  R.~Sato and K.~Tobioka,
  arXiv:1605.05366 [hep-ph];
  J.~M.~No,
  arXiv:1605.05900 [hep-ph];
  M.~A.~Ebert, S.~Liebler, I.~Moult, I.~W.~Stewart, F.~J.~Tackmann, K.~Tackmann and L.~Zeune,
  arXiv:1605.06114 [hep-ph];
  R.~Nevzorov and A.~W.~Thomas,
  arXiv:1605.07313 [hep-ph];
  K.~Kannike, G.~M.~Pelaggi, A.~Salvio and A.~Strumia,
  arXiv:1605.08681 [hep-ph];
  T.~Appelquist, J.~Ingoldby, M.~Piai and J.~Thompson,
  arXiv:1606.00865 [hep-ph];
  A.~Carmona, F.~Goertz and A.~Papaefstathiou,
  arXiv:1606.02716 [hep-ph].
  M.~Dalchenko, B.~Dutta, Y.~Gao, T.~Ghosh and T.~Kamon,
  arXiv:1606.03067 [hep-ph].
  C.~H.~Chen and T.~Nomura,
  arXiv:1606.03804 [hep-ph].
  L.~A.~Harland-Lang, V.~A.~Khoze, M.~G.~Ryskin and M.~Spannowsky,
  arXiv:1606.04902 [hep-ph].
  O.~Antipin, P.~Culjak, K.~Kumericki and I.~Picek,
  arXiv:1606.05163 [hep-ph].
  A.~Alves, A.~G.~Dias and K.~Sinha,
  arXiv:1606.06375 [hep-ph].
  M.~Carena, P.~Huang, A.~Ismail, I.~Low, N.~R.~Shah and C.~E.~M.~Wagner,
  arXiv:1606.06733 [hep-ph].
  A.~Efrati, J.~F.~Kamenik and Y.~Nir,
  arXiv:1606.07082 [hep-ph].
  C.~Royon,
  arXiv:1606.07675 [hep-ph].
  U.~K.~Dey, S.~Mohanty and G.~Tomar,
  arXiv:1606.07903 [hep-ph].
  A.~Di Iura, J.~Herrero-Garcia and D.~Meloni,
  arXiv:1606.08785 [hep-ph].
  S.~Banerjee, D.~Barducci, G.~Bà¥à¤à¤langer and C.~Delaunay,
  arXiv:1606.09013 [hep-ph].
  K.~Das, T.~Li, S.~Nandi and S.~K.~Rai,
  arXiv:1607.00810 [hep-ph].
  M.~Bauer, M.~Neubert and A.~Thamm,
  arXiv:1607.01016 [hep-ph].
  
\bibitem{not_higgs}
  D.~Becirevic, E.~Bertuzzo, O.~Sumensari and R.~Zukanovich Funchal,
  Phys.\ Lett.\ B {\bf 757}, 261 (2016);
  S.~Ghosh, A.~Kundu and S.~Ray,
  arXiv:1512.05786 [hep-ph];
  R.~Benbrik, C.~H.~Chen and T.~Nomura,
  Phys.\ Rev.\ D {\bf 93}, no. 5, 055034 (2016);
  X.~F.~Han and L.~Wang,
  Phys.\ Rev.\ D {\bf 93}, no. 5, 055027 (2016);
  W.~C.~Huang, Y.~L.~S.~Tsai and T.~C.~Yuan,
  Nucl.\ Phys.\ B {\bf 909}, 122 (2016);
  S.~Moretti and K.~Yagyu,
  Phys.\ Rev.\ D {\bf 93}, no. 5, 055043 (2016);
  M.~Badziak,
  Phys.\ Lett.\ B {\bf 759}, 464 (2016);
  K.~Cheung, P.~Ko, J.~S.~Lee, J.~Park and P.~Y.~Tseng,
  arXiv:1512.07853 [hep-ph];
  A.~Salvio and A.~Mazumdar,
  Phys.\ Lett.\ B {\bf 755}, 469 (2016);
  N.~Bizot, S.~Davidson, M.~Frigerio and J.-L.~Kneur,
  JHEP {\bf 1603}, 073 (2016);
  S.~K.~Kang and J.~Song,
  Phys.\ Rev.\ D {\bf 93}, no. 11, 115012 (2016);
  F.~Wang, W.~Wang, L.~Wu, J.~M.~Yang and M.~Zhang,
  arXiv:1512.08434 [hep-ph];
  F.~Wang, L.~Wu, J.~M.~Yang and M.~Zhang,
  Phys.\ Lett.\ B {\bf 759}, 191 (2016);
  X.~F.~Han, L.~Wang and J.~M.~Yang,
  Phys.\ Lett.\ B {\bf 757}, 537 (2016);
  C.~W.~Chiang and A.~L.~Kuo,
  arXiv:1601.06394 [hep-ph];
  M.~J.~Dolan, J.~L.~Hewett, M.~Kr\"amer and T.~G.~Rizzo,
  arXiv:1601.07208 [hep-ph];
  S.~Gopalakrishna and T.~S.~Mukherjee,
  arXiv:1604.05774 [hep-ph].
  
\bibitem{Dawson:2016ugw} 
  S.~Dawson and I.~M.~Lewis,
  arXiv:1605.04944 [hep-ph].  
  
  

\bibitem{eftfoundations}
  S.~Weinberg,
  Phys.\ Lett.\ B {\bf 91}, 51 (1980);
  S.~R.~Coleman, J.~Wess and B.~Zumino,
  Phys.\ Rev.\  {\bf 177}, 2239 (1969);
  C.~G.~Callan, Jr., S.~R.~Coleman, J.~Wess and B.~Zumino,
  Phys.\ Rev.\  {\bf 177}, 2247 (1969).

\bibitem{eftorig}
  C.~J.~C.~Burges and H.~J.~Schnitzer,
  Nucl.\ Phys.\ B {\bf 228}, 464 (1983); 
  C.~N.~Leung, S.~T.~Love and S.~Rao,
  Z.\ Phys.\ C {\bf 31}, 433 (1986); 
  W.~Buchm\"uller and D.~Wyler,
  Nucl.\ Phys.\ B {\bf 268}, 621 (1986);
  W.~Kilian,
  Springer Tracts Mod.\ Phys.\  {\bf 198}, 1 (2003).
  A.~De~Rujula, M.~Gavela, P.~Hernandez, and E.~Masso,
  \newblock Nucl.Phys. {\bf B384}, 3 (1992);
  K.~Hagiwara, R.~Szalapski, and D.~Zeppenfeld,
  Phys.Lett. {\bf B318}, 155 (1993);
  K.~Hagiwara, S.~Matsumoto, and R.~Szalapski,
  Phys.Lett. {\bf B357}, 411 (1995);
  M.~C.~Gonzalez-Garcia,
  Int.J.Mod.Phys. {\bf A14}, 3121 (1999);
  G.~Passarino,
  Nucl.\ Phys.\ B {\bf 868}, 416 (2013).
  K.~Hagiwara, S.~Ishihara, R.~Szalapski, and D.~Zeppenfeld,
  Phys.Rev. {\bf D48}, 2182 (1993);
  K.~Hagiwara, T.~Hatsukano, S.~Ishihara and R.~Szalapski,
  Nucl.\ Phys.\ B {\bf 496}, 66 (1997);
  R.~Alonso, M.~B.~Gavela, L.~Merlo, S.~Rigolin and J.~Yepes,
  Phys.\ Lett.\ B {\bf 722}, 330 (2013)
  Erratum: [Phys.\ Lett.\ B {\bf 726}, 926 (2013)];
  G.~Buchalla, O.~Cat\`a and C.~Krause,
  Nucl.\ Phys.\ B {\bf 880}, 552 (2014);
  M.~B.~Gavela, J.~Gonzalez-Fraile, M.~C.~Gonzalez-Garcia, L.~Merlo, S.~Rigolin and J.~Yepes,
  JHEP {\bf 1410}, 44 (2014);
  
  \bibitem{higgsreview}
  C.~Englert, A.~Freitas, M.~M.~M\"uhlleitner, T.~Plehn, M.~Rauch, M.~Spira and K.~Walz,
  J.\ Phys.\ G {\bf 41}, 113001 (2014).
  
  \bibitem{Grzadkowski:2010es} 
  B.~Grzadkowski, M.~Iskrzynski, M.~Misiak, and J.~Rosiek,
  JHEP {\bf 1010}, 085 (2010);
  
  

\bibitem{Gripaios:2016xuo} 
  B.~Gripaios and D.~Sutherland,
  arXiv:1604.07365 [hep-ph].
  
 
\bibitem{Kamenik:2016tuv} 
  J.~F.~Kamenik, B.~R.~Safdi, Y.~Soreq and J.~Zupan,
  arXiv:1603.06566 [hep-ph].
  
\bibitem{Bauer:2016lbe}
  M.~Bauer, C.~Hoerner and M.~Neubert,
  arXiv:1603.05978 [hep-ph].
  
\bibitem{Franceschini:2016gxv} 
  R.~Franceschini, G.~F.~Giudice, J.~F.~Kamenik, M.~McCullough, F.~Riva, A.~Strumia and R.~Torre,
  arXiv:1604.06446 [hep-ph].
  
  
  
		

\bibitem{sfitter_higgs}
 R.~Lafaye, T.~Plehn, M.~Rauch, D.~Zerwas and M.~D\"uhrssen,
  JHEP {\bf 0908}, 009 (2009);
   M.~Klute, R.~Lafaye, T.~Plehn, M.~Rauch and D.~Zerwas,
  Phys.\ Rev.\ Lett.\  {\bf 109}, 101801 (2012);
 T.~Plehn and M.~Rauch,
  Europhys.\ Lett.\  {\bf 100}, 11002 (2012).

\bibitem{sfitter_susy}
  R.~Lafaye, T.~Plehn, M.~Rauch and D.~Zerwas,
  Eur.\ Phys.\ J.\ C {\bf 54}, 617 (2008).

\bibitem{rfit}
 A.~H\"ocker, H.~Lacker, S.~Laplace and F.~Le Diberder,
  Eur.\ Phys.\ J.\  C {\bf 21}, 225 (2001).
  

\bibitem{S750_2000} 
  V.~Khachatryan {\it et al.} [CMS Collaboration],
  Phys.\ Lett.\ B {\bf 750}, 494 (2015)

\bibitem{S750_1010}
  G.~Aad {\it et al.} [ATLAS Collaboration],
  JHEP {\bf 1601}, 032 (2016)

\bibitem{S750_1110}
  The ATLAS collaboration,
  ATLAS-CONF-2016-021.

\bibitem{S750_1020}
  G.~Aad {\it et al.} [ATLAS Collaboration],
  Eur.\ Phys.\ J.\ C {\bf 76}, no. 1, 45 (2016)

\bibitem{S750_1120}
  The ATLAS collaboration,
  ATLAS-CONF-2016-016.

\bibitem{S750_1121}
  The ATLAS collaboration,
  ATLAS-CONF-2016-012.

\bibitem{S750_1130}
  The ATLAS collaboration,
  ATLAS-CONF-2016-010.

\bibitem{S750_2130}
  CMS Collaboration [CMS Collaboration],
  CMS-PAS-EXO-16-019.

\bibitem{S750_1030}
  G.~Aad {\it et al.} [ATLAS Collaboration],
  Phys.\ Lett.\ B {\bf 738}, 428 (2014)

\bibitem{S750_1040}
  G.~Aad {\it et al.} [ATLAS Collaboration],
  JHEP {\bf 1508}, 148 (2015)

\bibitem{S750_2050}
  V.~Khachatryan {\it et al.} [CMS Collaboration],
  arXiv:1604.08907 [hep-ex].

\bibitem{S750_1160}
  The ATLAS collaboration,
  ATLAS-CONF-2016-017.

\bibitem{S750_2060}
  V.~Khachatryan {\it et al.} [CMS Collaboration],
  Phys.\ Lett.\ B {\bf 749}, 560 (2015)

\bibitem{S750_2070}
  CMS Collaboration, 
  CMS-PAS-HIG-14-028.

\bibitem{Goertz:2014qia} 
  F.~Goertz,
  Phys.\ Rev.\ Lett.\  {\bf 113}, no. 26, 261803 (2014).

\bibitem{tim}
  S.~Liem, G.~Bertone, F.~Calore, R.~R.~de Austri, T.~M.~P.~Tait, R.~Trotta and C.~Weniger,
  arXiv:1603.05994 [hep-ph].
  
  
\end{thebibliography}
\end{document}